\documentclass[12pt,a4paper,final]{iopart}

\usepackage{amsmath,amsfonts,amssymb}
\usepackage{iopams}  
\usepackage[utf8]{inputenc}

\usepackage{graphicx}

\usepackage[english]{babel}
\usepackage{upgreek}

\usepackage{bm}

\usepackage{textcomp} 
\usepackage[autolanguage]{numprint}
\usepackage{expl3,siunitx}
\ExplSyntaxOn
\cs_new_eq:NN \fpcmpTF \fp_compare:nTF
\ExplSyntaxOff

\usepackage{tabulary}
\begin{document}
	
	\title[The Gravitational Wave Observatory Designer]{The Gravitational Wave Observatory Designer: Sensitivity Limits of Spaceborne Detectors}
	
	\author{S Barke$^{1}$, Y Wang$^{1,2}$, JJ Esteban Delgado$^{1,3}$, M Tröbs$^{1}$, G Heinzel$^{1}$, K Danzmann$^{1}$}
	\address{$^1$Max Planck Institute for Gravitational Physics (Albert-Einstein-Institut) and Leibniz Universität Hannover, Callinstr. 38, D-30167 Hannover, Germany}
	\address{$^2$present address: The University of Western Australia, School of Physics, 35 Stirling Highway, Crawley, Perth, Western Australia 6009}
	\address{$^3$present address: Coherent LaserSystems GmbH \& Co. KG, Garbsener Landstraße 10, D-30419 Hannover, Germany}
	\ead{simon.barke@aei.mpg.de}

	\begin{abstract}
		The most promising concept for low frequency gravitational wave observatories are laser interferometric detectors in space.  It is usually assumed that the noise floor for such a detector is dominated by optical shot noise in the signal readout. For this to be true, a careful balance of mission parameters is crucial to keep all other parasitic disturbances below shot noise. We developed a web application that uses over 30 input parameters and considers many important technical noise sources and noise suppression techniques. It optimizes free parameters automatically and generates a detailed report on all individual noise contributions. Thus you can easily explore the entire parameter space and design a realistic gravitational wave observatory.
		
		In this document we describe the different parameters, present all underlying calculations, and compare the final observatory's sensitivity with astrophysical sources of gravitational waves. We use as an example parameters currently assumed to be likely applied to a space mission to be launched in 2034 by the European Space Agency. The web application itself is publicly available on the Internet at http://spacegravity.org/designer.
		
	\end{abstract}
	
	\submitto{\CQG}
	
	\section{Introduction}

	\begin{figure}[htb!]\centering 
		\includegraphics[width=\linewidth]{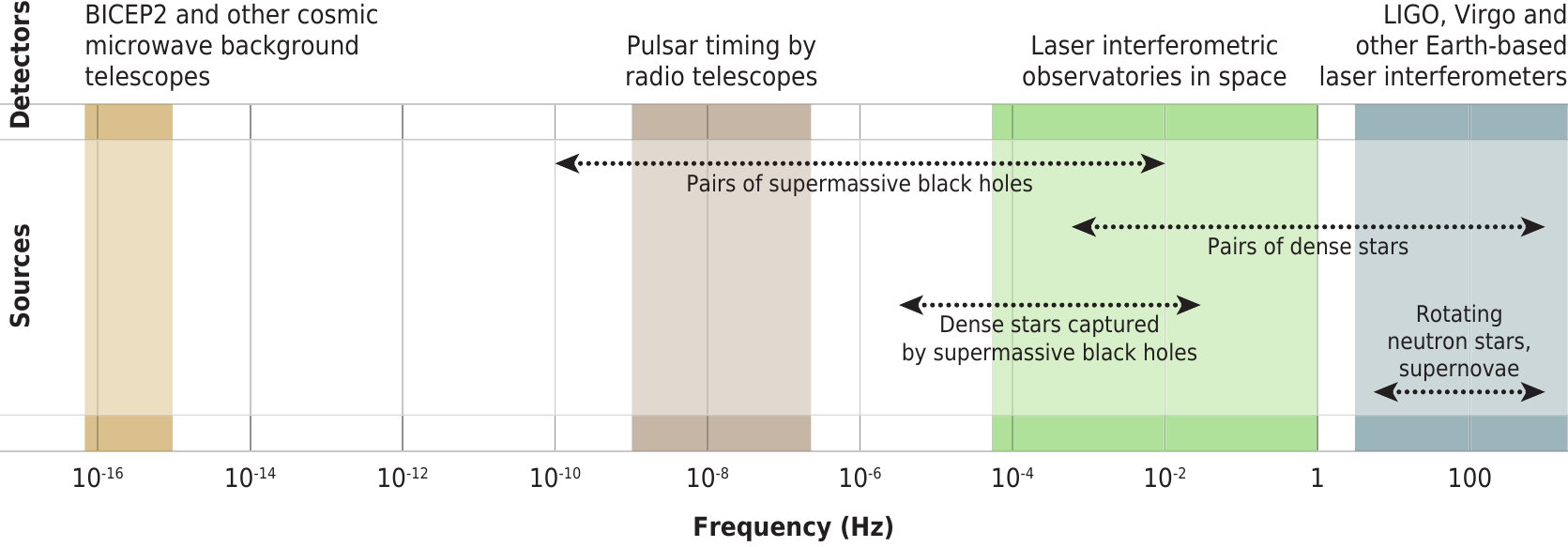}
		\caption{Frequency range of gravitational wave sources and bandwidth of corresponding gravitational wave detectors on Earth and in space. A gravitational wave background generated during cosmic inflation should be present over the entire frequency spectrum.}
		\label{fig:view}
	\end{figure}

	Gravitational waves \cite{einstein1937gravitational} are expected to be the next big revelation in astronomy, cosmology, and fundamental physics alike. In contrast to electromagnetic radiation, gravitational radiation travels unimpeded throughout the entire universe, and even electromagnetically dark objects are capable of producing gravitational waves. Their continuous observation will enable us to study these dark objects directly for the very first time.\\
	
	Alongside indirect yet irrefutable proof of the existence of gravitational waves \cite{taylor1989further}, research teams also look into evidence for gravitational waves produced during cosmic inflation, now red-shifted to a static polarization pattern imprint in the cosmic microwave background radiation \cite{ade2014bicep2, mortonson2014joint}. But there are many other sources out there: Very low frequency gravitational waves below 1\,\textmu Hz produced by pairs of supermassive black holes can be detected when timing millisecond pulsars with radio telescopes \cite{0264-9381-27-8-084013}.
	High frequency gravitational waves above 10\,Hz---as produced by rotating neutron stars or asymmetric supernovae---will be measured by sophisticated Earth-based laser interferometric detectors \cite{harry2010advanced, accadia2011status, somiya2012detector, grote2008status}. 
	Some of the most interesting sources of gravitational waves (like supermassive black hole mergers, dense stars captured by supermassive black holes, and pairs of dense stars) emit at frequencies between 10\,\textmu Hz and 10\,Hz, see Figure~\ref{fig:view}. However, due to seismic disturbances and environmental gravity variations, this frequency range is not accessible from Earth. 
	Hence a spaceborne gravitational wave observatory was recently selected by the European Space Agency (ESA) to be launched in the 2030s as 3rd large mission of the Cosmic Vision program. Laser interferometric detectors \cite{seoane2013gravitational} are generally considered to be the most promising option for the intended purpose.
	
	Concepts of such interferometric observatories feature multiple spacecraft separated by millions of kilometers that form a giant laser interferometer, compare Laser Interferometer Space Antenna (LISA) \cite{danzmann2011lisa}, New Gravitational wave Observatory (NGO) \cite{jenrich2012ngo}. 
	Usually documents refer to one of these carefully thought out design studies and determine the observatory's sensitivity by just three parameters: the well known interferometer topology, its optical shot noise limit, and acceleration noise of gravitational reference points (proof masses). 
	When one starts exploring the broader parameter space with regard to ESA's Cosmic Vision mission, it might become impossible to keep all technical noise sources below the interferometer's shot noise within the limits of current technology. This document will take the reader through each step of the design process, explain the influences of design choices on the observatory's sensitivity, and point out potential limitations. This will help you to carefully balance out all mission parameters in the associated web application where you can design a realistic gravitational wave observatory with your very own set of parameters.
	
	\section{Mission Parameters}
	
\begin{figure}[htb!]\centering
		\includegraphics[width=.675\linewidth]{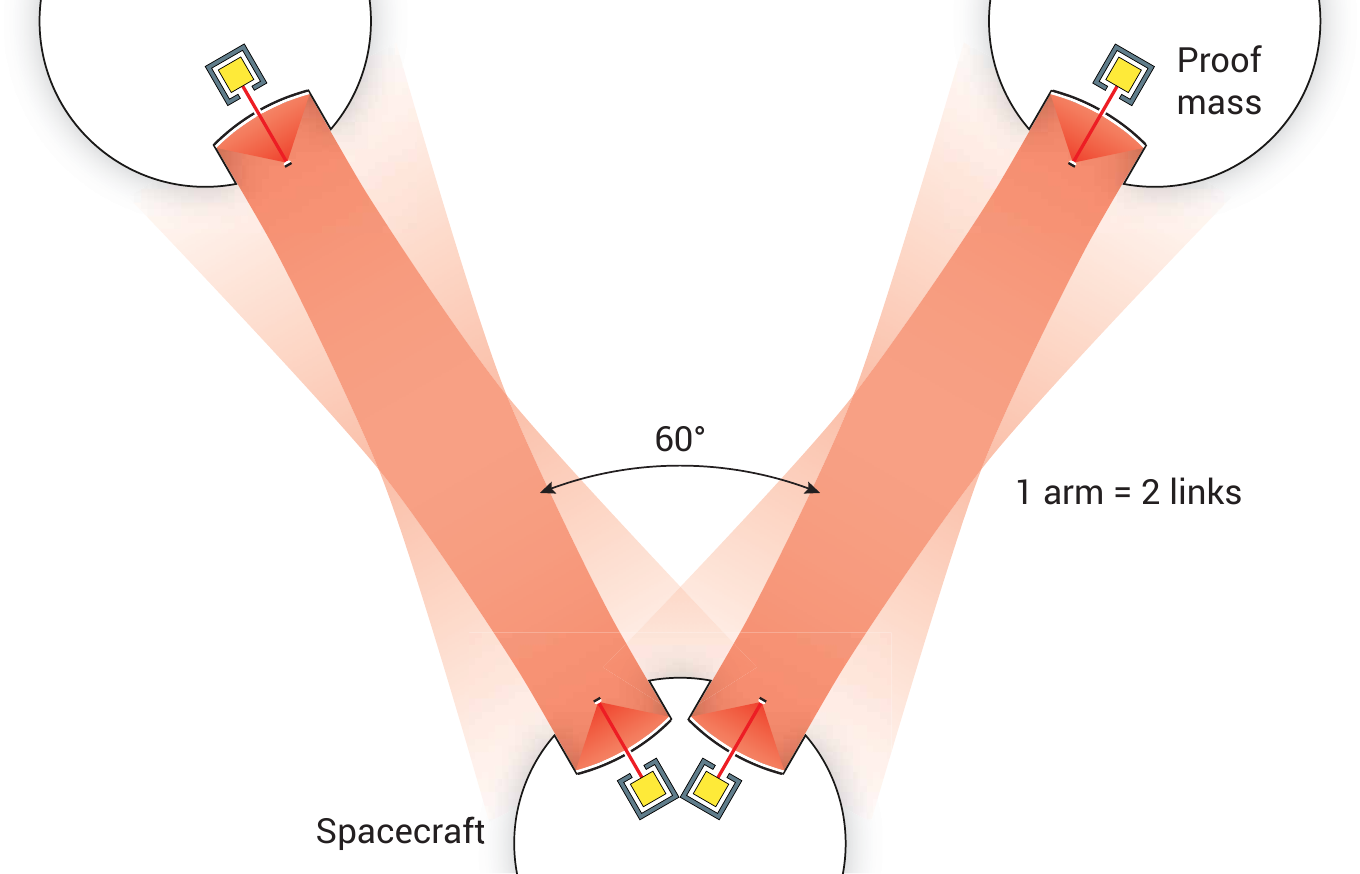}
		\caption{A laser interferometric gravitational wave observatory in space consists of a minimum of three spacecraft that form a virtual two-arm Michelson interferometer with four individual laser links. Freely floating proof masses act as gravitational reference points.}
		\label{fig:virtual-michelson}
	\end{figure}

	A laser interferometric gravitational wave observatory in space consists of a virtual Michelson interferometer that measures changes in the proper distance between gravitational reference points: freely floating proof masses that form the end mirrors of the interferometer arms. This concept is illustrated in Figure~\ref{fig:virtual-michelson}. Gravitational waves will alter this distance in different proportions for the individual arms depending on their polarization and sky position.\\

\begin{figure}[htb!]\centering 
			\includegraphics[width=\linewidth]{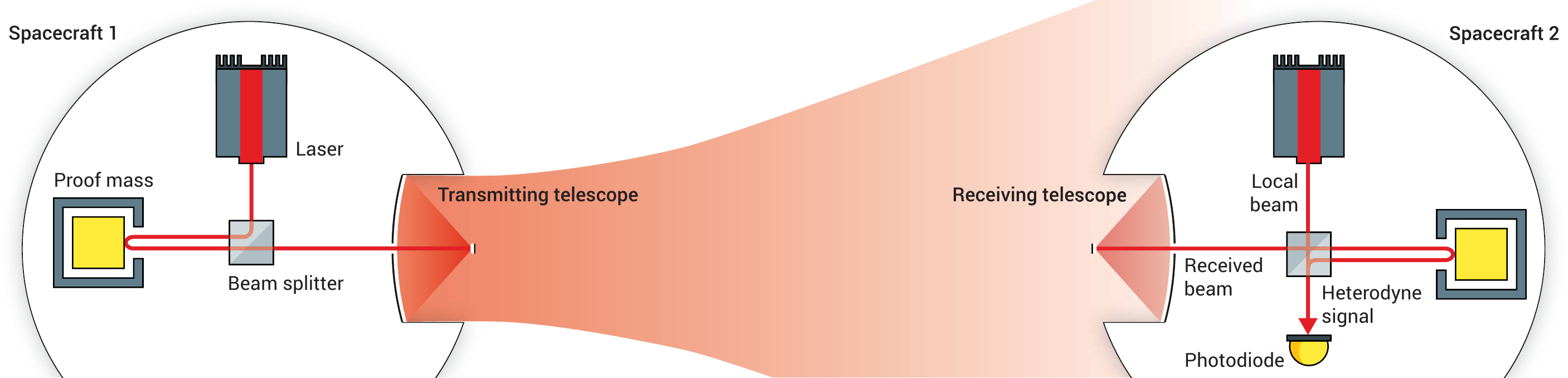}
			\caption{Simplistic illustration of one individual laser link between two spacecraft of a spaceborne gravitational wave observatory. A remote laser (on Spacecraft 1) is transmitted to Spacecraft 2 via optical telescopes. Here it gets interfered with a local laser of different frequency and the heterodyne signal is detected by a photodiode. Gravitational waves alter the proper distance between the spacecraft resulting in a phase shift of the heterodyne signal. Freely floating proof masses form the end points of the inter-spacecraft interferometer arm to suppress the influence of spacecraft position jitter on the actual arm length. To construct a complete observatory arm, you also need the reverse link that transmits light from Spacecraft 2 to Spacecraft 1.}
			\label{fig:singlelink}
		\end{figure}

	The virtual Michelson interferometer is constructed from individual `links', each individual link consists of one or more actual laser interferometers, see Figure~\ref{fig:singlelink}. It  will use a heterodyne detection scheme that interferes laser light from a distant spacecraft (received beam) with an on-board laser (local beam) at a recombination beam splitter. Optical pathlength fluctuations between proof masses will shift the phase of the received beam. These phase shifts are conserved in the heterodyne process, thus the phase of the heterodyne signal contains the gravitational wave signal. One observatory arm always consists of two counterpropagating links.

	\begin{figure}[htb!]\centering
		\includegraphics[width=.675\linewidth]{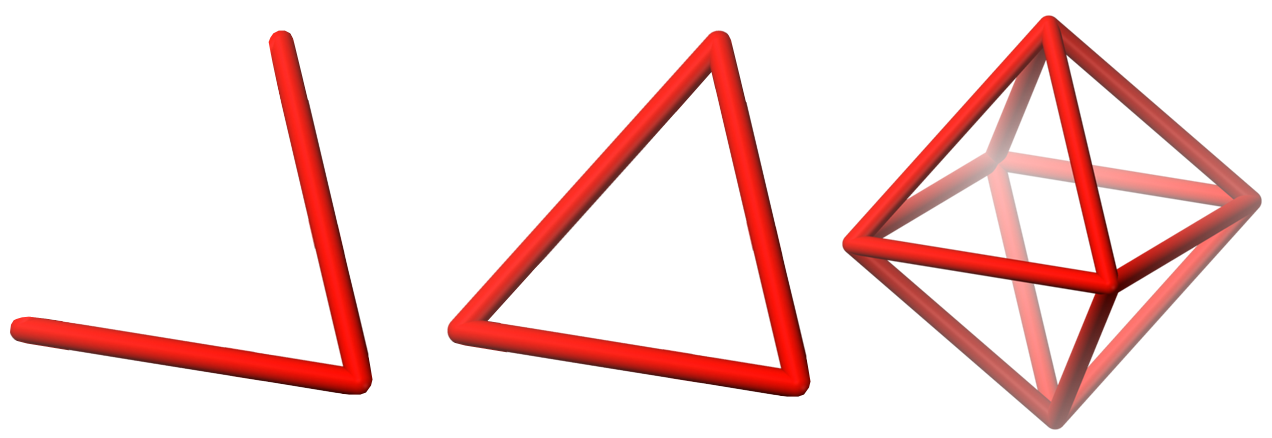}
		\caption{Possible arrangements for interferometric gravitational wave observatories: two-arm (left), triangular (center), octahedral (right) -- corner points mark the position of the individual spacecraft.}
		\label{fig:configuration}
	\end{figure}

		While a minimum of two arms (four laser links) between three spacecraft is required to construct the virtual interferometer, more links will not only improve the observatory's sensitivity but also produce other consequential benefits:
		A triangular three-arm (6 link) detector can discriminate between different gravitational wave polarizations instantaneously and yields a much better spatial resolution. An octahedral 12-arm (24 link) observatory \cite{wang2013octahedron} would in theory be able to suppress acceleration noise on the proof masses alongside other else limiting noise sources. 
		Possible arrangements are shown in Figure~\ref{fig:configuration}.

	\subsection{Constellation}\label{sec:constellation}

Beside the number of arms, there are other fundamental design choices that determine the capabilities of your observatory.

\paragraph{Arm length}
The most consequential mission parameter is the separation between spacecraft that resembles the arm length of the virtual interferometer. It has multiple effects on the observatory's sensitivity. Longer arms make it more sensitive to lower gravitational wave frequencies but also decrease the received laser light power thus increasing the amount of shot noise in the signal. Also the arm length has an impact on orbit stability.
The gravitational wave sources commonly targeted by spaceborne observatories are in the millihertz range with wavelengths of $10^9$\,km and more, consequently the optimal arm length should be on the order of million kilometers. Even the observation of gravitational waves at hertz with cycle durations of the order of seconds still requires arm lengths of some thousand kilometers.


\paragraph{Orbit}
Drifts in the spacecraft constellation result in Doppler shifts of the laser light. Hence the interferometer requires a readout that measures the phase of a high frequency heterodyne signal. A lower heterodyne frequency simplifies the phase readout.
Switchable offset frequency phase-locked loops between lasers minimize the maximum heterodyne frequency. At the same time they avoid zero crossings and other forbidden frequency domains. The effectiveness of this effort is limited by the orbit stability and the resulting magnitude of the Doppler shifts.

For a triangular constellation with average arm lengths of \numprint{5000000}\,km in a heliocentric orbit $20^\circ$ behind Earth studies predict a heterodyne frequency of less than 25\,MHz \cite{SBarkePhD}. A smaller separation in heliocentric orbits would further reduce this value by some megahertz per \numprint{1000000}\,km arm length. The stability of geocentric orbits would greatly suffer from the proximity to the Earth-Moon system, amplifying the Doppler shifts and increasing the maximum heterodyne frequency. For octahedral (24 link) constellations, so far only short arm ($< 1500$\,km) halo orbits near the Lagrangian point L1 have been found to be stable enough, with Doppler shifts being still under investigation.\\

The general feasibility of a chosen constellation with a specific spacecraft separation in a certain orbit must be subject to a more detailed study which in turn will reveal the time-varying Doppler shifts. 
A customized laser locking scheme and frequency swapping plan that considers a wide variety of auxiliary functions \cite{heinzel2011auxiliary} and technical limitations then sets the maximum heterodyne frequency. In the following we will work with a  triangular three-arm ($N_\text{links} = 6$ links) formation  featuring a reasonable arm length of $L_\text{arm} = \numprint{2000000}$\,km in a heliocentric orbit so that a maximum heterodyne frequency of $f_\text{het} = 18$\,MHz can be assumed.

\subsection{Lasers, Optics, and Photoreceivers}\label{sec:optelec}

To decrease the read-out noise level of your observatory, it is not only beneficial to have high-quality photoreceivers but also to increase and stabilize the laser power received by the remote spacecraft (see Section~\ref{ch:readoutnoise}). For the power increase you can shrink down the arm length (which has an adverse effect on the overall sensitivity) or increase the laser power (which will result in a higher power consumption) and enlarge the optical telescopes (which increases the size of the spacecraft and thereby the mission cost). Balancing these parameters within the mission's financial constraints is crucial.

\paragraph{Lasers} All lasers have to meet certain stability requirements. Fluctuations in the laser power relative to the average absolute power level, the so-called relative intensity noise (RIN), will directly couple to the photocurrent of the receiving photo detector as one part of the read-out noise and deteriorate the interferometric length measurements. The best space qualified lasers available as of this writing meet a relative intensity noise of $RIN = \num{1e-8}$\,$/\sqrt{\text{Hz}}$ for Fourier frequencies above 5\,MHz at $\lambda_\text{laser} = \num{1064}$\,nm\footnote{$1064$\,nm is a standard wavelength for gravitational wave observatories. At other wavelengths relative intensity noise and frequency noise might be very different. There are additional consequences: While phase noise would have a smaller impact on the displacement noise at shorter wavelength (see Equation~(\ref{eq:radinpm})), instabilities in the spacecraft orbits would result in higher Doppler shifts and hence increase the maximum heterodyne frequency.} wavelength. Below this frequency the noise increases significantly so that no measurements at heterodyne frequencies below 5\,MHz are possible. This limitation determines a forbidden domain for the frequency swapping plan mentioned in Section~\ref{sec:constellation}. For other relative intensity noise levels this lower frequency might be different.
	
	Frequency noise of the lasers will couple via the arm length difference of individual interferometers into phase fluctuations in the signal read-out. That is why one master laser is pre-stabilized by a reference cavity, a molecular frequency standard or similar techniques, and all other lasers will be actively locked onto this master laser. The residual frequency noise after pre-stabilization is assumed to be $\widetilde{\vartheta}_\text{pre} = \num{290}$\,$\text{Hz}/{\sqrt{\text{Hz}}}$. To simplify measures, this noise contribution---like most within in this document---is given as white noise valid at the targeted gravitational wave frequency range.
	
	The laser power---or, more importantly, the power passed to the transmitting telescope---possibly depends not only on the actual master laser but also on a laser amplifier. The above values for relative intensity noise and frequency noise after pre-stabilization already consider the presence of such an amplifier stage. In the following we consider a power passed to the transmitting telescope $P_\text{tel}=\num{1.65}$\,W.
	
	\paragraph{Received Laser Power}
	
	For the amount of light transmitted between spacecraft, the telescope diameter is the important parameter. In the following, we will assume a telescope with a moderate $ d_\text{tel} = \num{26}$\,cm diameter primary mirror. We can now calculate the laser power received by the remote spacecraft. There are three different cases: 
	
	\begin{enumerate} 
		\item \textbf{Short arms / big telescope mirrors}, where the full Gaussian beam fits well within the telescope when the waist is located at the center between the spacecraft. Here we can transmit the full laser power.
		\item \textbf{Long arms / small telescope mirrors}, where the Gaussian beam has expanded to a width much larger than the receiving telescope when the waist is located at the telescope aperture. Here we cut out a `flat-top' beam out of a field of constant intensity.
		\item \textbf{Anything in between}, where the Gaussian beam is larger than the telescope diameter but too small for a flat intensity profile. This case should be avoided since the received power will be subject to beam pointing, a property that is not considered by the web application.
	\end{enumerate}
	
	To check if we can  transmit the full laser power by setting the waist of the beam at the center between the spacecraft separated by $L_\text{arm}$, we compute the optimum waist radius $\omega_0$ for a minimum Gaussian beam radius $\omega(x)$ at $x=L_\text{arm}/2$ apart from the waist:
	\begin{equation}
	\omega(x) = \omega_0 \times \sqrt{1+\left(\frac{x \times \lambda_\text{laser}}{\pi \omega_0^2}\right)^2}~.
	\label{eq:waist}
	\end{equation}
	
	For an arm length of \numprint{2000000}\,km, the optimum waist is found to be \num{18.40}\,m and the observatory would require telescopes with a diameter lager than \num{50}\,m to transmit the full laser power. Thus we abandon this plan and intend to optimize the beam parameters for a maximum light intensity across the receiving telescope. As deduced from \cite{holmes1970axis} the maximum intensity is reached for the waist placed at the transmitting telescope's aperture. For long arms the on-axis far-field intensity at the receiver can then be expressed as
	\begin{equation}
	I_\text{rec} = \frac{\pi\: P_\text{tel}\: d_\text{tel}^2}{2\:L_\text{arm}^2 \: \lambda_\text{laser}^2} \times \underbrace{\alpha^2 e^{-\frac{2}{\alpha^2}}\left( e^\frac{1}{\alpha^2}-1\right)^2}_{\text{max}() = 0.4073 ~ \text{for} ~ \alpha = 0.8921}  ~,
	\label{eq:intensity}
	\end{equation}
	where $\alpha$ is the waist radius in units of the telescope radius: $\omega_0 = \alpha \times d_\text{tel}/2 \: $. The maximum of this function occurs at $\alpha = 0.8921$ as indicated above, so that the optimum waist $\omega_0 = 0.8921 \times d_\text{tel}/2 = 11.6$\,cm. Accordingly the best achievable intensity at the receiver is $I_\text{rec} = 15.76$\,$\text{nW}/\text{m}^2$.
	
	If we use a smaller beam that completely passes through the telescope, its divergence would be larger and the beam would be spread over a bigger area at the receiver so the intensity would be  smaller. If we use a larger beam with a smaller divergence, a larger fraction of the beam power would be rejected by the transmitting telescope aperture and again the intensity at the receiver would be smaller. In the above equation, diffraction effects for the beam truncated by a circular aperture were taken into account. The off-axis intensity distribution shows some curvature and diffraction rings, so that strictly speaking one cannot state a Gaussian beam radius. Following Equation~(\ref{eq:waist}) to get an approximate far end beam diameter, we obtain $d_\text{rec} = 2\times \omega\left(L_\text{arm}\right) = 11.68$\,km. This is much larger than the telescope diameter and we can confidently assume a flat intensity profile.\\
	
	
	The received laser power now easily results from the light intensity at the receiving telescope multiplied with its optical efficiency and the collection area,
	\begin{equation}
	P_\text{rec} = \pi\:\left(\frac{d_\text{tel}}{2}\right)^2 \eta_\text{opt} \: I_\text{rec} = \num{836.61}\,\text{pW}~.
	\label{eq:recpower}
	\end{equation}
	Here $\eta_\text{opt} =  \num{70}$\,\% denotes an overall optical efficiency in the receive pass that  accounts for all losses in the optical path from the transmitting telescope to the recombination beam splitter on the receiving spacecraft.
	
	\paragraph{Optical Bench}\label{sec:opticalbench}
	
	Interferometers are used to optically read out the displacement of the proof masses. These interferometers are constructed with fused silica optics that are bonded to an optical bench made out of an ultra-low expansion glass-ceramic. There are different possible interferometer topologies. In principle the simplistic scheme illustrated in Figure~\ref{fig:singlelink} would suffice since the difference of the two heterodyne signals (both links) cancels not only noise induced by the laser feeds (optical fibers from the laser to the optical bench) but also spacecraft position noise and even phase noise caused by temperature fluctuations of the optical bench. At the same time changes in the proper distance between the spacecraft (including gravitational waves) are preserved.
	
\begin{figure}[htb!]\centering
		\includegraphics[width=.675\linewidth]{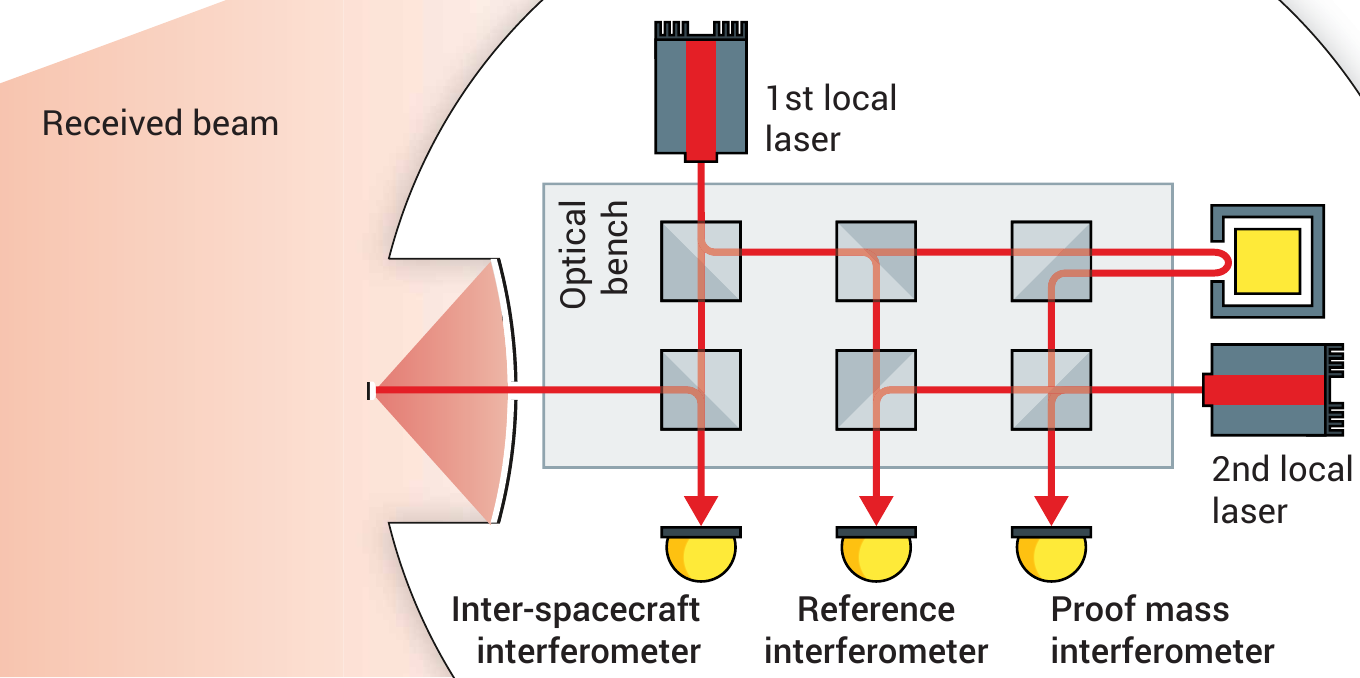}
		\caption{The measurement of the proper distance between any two proof masses is split into individual interferometers. Here, at each end of the link there are three interferometers, one to read out the inter-spacecraft distance, one to determine the displacement of the local proof mass in relation to the optical bench, and one acting as a reference.
		}
		\label{fig:opd}
	\end{figure}

	More complex topologies exist that split the single link measurement into smaller sections that are read out by individual interferometers \cite{jennrich2009lisa}. For example one could omit the reflection of the received beam on the local proof mass. Instead, the proof mass displacement would then be determined with respect to the optical bench with a dedicated proof mass interferometer. This simplifies integration and testing of the interferometers and allows for easier beam alignment. Observatories that receive only low optical power from the remote spacecraft benefit from a scheme with three interferometers as illustrated in Figure~\ref{fig:opd}. Here, a second local laser is used in the proof mass interferometer so that the full power of the received beam can be utilized in an inter-spacecraft interferometer. This scheme requires an additional reference interferometer  to cancel the noise induced by the laser feeds.\\
	
	The heterodyne signal with the lowest amplitude (and thus possibly a limiting factor) usually is the one of the inter-spacecraft interferometer. Here, the heterodyne efficiency at the recombination beam splitter---a factor describing the mode overlap between the two laser beams---gains importance. It is assumed to be $\eta_\text{het} = \num{70}$\,\%. A higher efficiency increases the signal that is received by the photo detector.
	
	\paragraph{Photoreceivers} 
	
	
	The heterodyne signal from the recombination beam splitter  is detected by a photodiode. A transimpedance amplifier converts the photocurrent into a proportional voltage. The quantum efficiency of the photo detector is assumed to be $\eta_\text{pd} = \num{80}$\,\%\footnote{$\num{80}$\,\% is a typical quantum efficiency for InGaAs photodiodes at \num{1064}\,nm.}. This translates to a photodiode responsivity of
	\begin{equation}
	R_\text{pd} = \eta_\text{pd}\:\frac{q_\text{e}\:\lambda_\text{laser}}{h\:c} = \num{0.69}\,\frac{\text{A}}{\text{W}}
	\label{eq:response}
	\end{equation}
	where $q_\text{e}$ is the electron charge, $h$ is Planck's constant, and $c$ is the speed of light.
	
	The signal quality depends on the current noise of the amplifier, which consists of the input current noise ($\widetilde{I}_\text{pd}$, set to $\num{2}$\,$\text{pA}/{\sqrt{\text{Hz}}}$) and intrinsic voltage noise of the amplifier ($\widetilde{U}_\text{pd}$, set to $\num{2}$\,$\text{nV}/{\sqrt{\text{Hz}}}$) that is converted to current noise by the impedance of the photodiode. With an assumed photodiode capacitance $C_\text{pd}=10$\,pF this impedance is given by 
	\begin{equation}
	Z_\text{pd} = \frac{1}{2\pi\:C_\text{pd}\:f_\text{het}} = \num{884.22}\, \Omega ~.
	\label{eq:impedance}
	\end{equation}
	The higher the heterodyne frequency $f_\text{het}$ or capacitance, the lower the impedance becomes, which in turn will increase the resulting current noise of the transimpedance amplifier.
	
\begin{figure}[htb!]\centering
		\includegraphics[width=.675\linewidth]{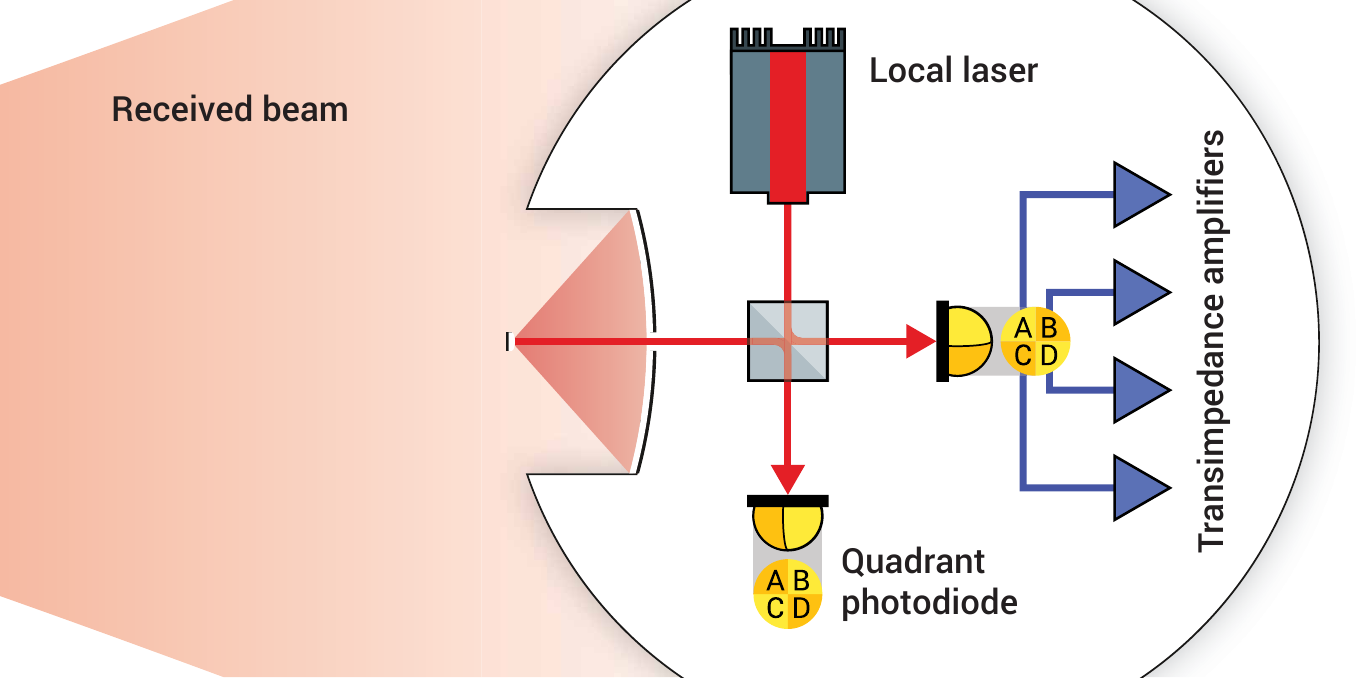}
		\caption{Two quadrant photodiodes (four segments each, one at each output port of a 50:50 beam splitter) are used to read out the heterodyne signal. Each segment is connected to a transimpedance amplifier.}
		\label{fig:pd}
	\end{figure}

	The various noise quantities in the signal add up differently depending on the number of photodiode segments used in the detection. In this study we consider one pair of redundant quadrant photodiodes with four segments each ($N_\text{pd} = \num{4}$) as illustrated in Figure~\ref{fig:pd}.
	
\subsection{Temperature Stability}\label{sec:temperature}

\begin{figure}[htb!]\centering
	\includegraphics[width=.675\linewidth]{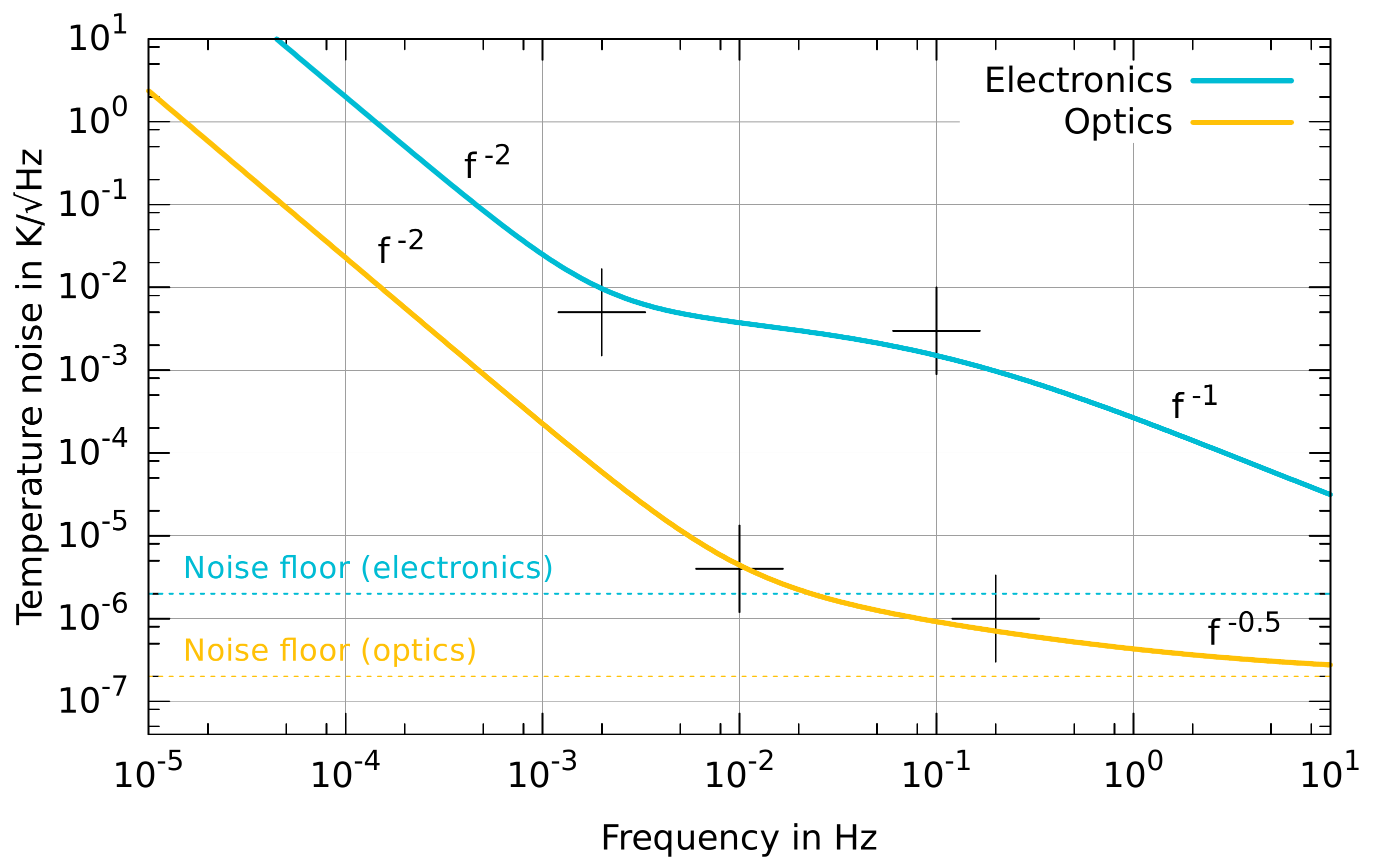}
	\caption{Temperature noise at the electronics and electro-optics (blue) and at the optical bench (yellow) in the significant heterodyne frequency range.}
	\label{fig:temperature-noise}
\end{figure}

Some components will shift the overall optical path length or in general the phase of essential signals when a change in temperature occurs. While we assume a constant path length noise over the measurement band for the telescope (see Section~\ref{ch:pathlengthnoise}) we will use a more complex temperature noise model to calculate the influence on the optical bench as well as some electronic and electro-optical components. Figure~\ref{fig:temperature-noise} shows a plot of the assumed temperature noise in Kelvin$/\sqrt{\text{Hz}}$ over Fourier frequency $f$. The web application allows you to set a noise floor, two corner frequencies and a lower and upper slope for each noise model.

The blue trace corresponds to the temperature noise at the electronics and electro-optics, $\widetilde{T}_\text{el}\left(f\right)$, usually distributed in boxes within the spacecraft, and features noise levels of $\num{5}$\,$\text{mK}/{\sqrt{\text{Hz}}}$ and $\num{3}$\,$\text{mK}/{\sqrt{\text{Hz}}}$ at $f = \num{2e-3}$\,Hz and $f = \num{1e-1}$\,Hz respectively. The slopes below and above these corner frequencies are $f^{\num{-2}}$ and $f^{\num{-1}}$ with a constant noise floor of $\num{0.002}$\,$\text{mK}/{\sqrt{\text{Hz}}}$. The yellow trace corresponds to the temperature noise at the optical bench, $\widetilde{T}_\text{ob}\left(f\right)$, which is placed at the center of the spacecraft where the temperature is commonly more stable. We assume noise levels of $\num{0.004}$\,$\text{mK}/{\sqrt{\text{Hz}}}$ and $\num{0.001}$\,$\text{mK}/{\sqrt{\text{Hz}}}$ at $f = \num{1e-2}$\,Hz and $f = \num{2e-1}$\,Hz respectively. The slopes below and above these corner frequencies are $f^{\num{-2}}$ and $f^{-0.5}$ with a constant noise floor of $\num{0.0002}$\,$\text{mK}/{\sqrt{\text{Hz}}}$.

\section{Displacement Noise Contributions}

While we will read out phase shifts $\delta \phi$ in the heterodyne signal, a more intuitive quantity is the apparent spacecraft displacement, $\delta x$, that results from a measured phase shift. Since phase shifts in the individual laser beams are preserved in the heterodyne signal, the conversion between both is expressed by
\begin{equation}
\delta x = \frac{\lambda_\text{laser}}{2\pi} \times \delta \phi ~ .
\label{eq:radinpm}
\end{equation}
The same is true for the conversion between linear spectral densities of phase noise $\widetilde{\phi}$ (given in rad$/\sqrt{\text{Hz}}$) and displacement noise $\widetilde{x}$ (given in m$/\sqrt{\text{Hz}}$) and used throughout this document.\\

There are multiple noise sources that are indistinguishable from an actual spacecraft displacement due to gravitational waves, any one of which could in principle limit the observatory's sensitivity. In the following we will compute each displacement noise contribution individually.

\subsection{Read-out Noise}\label{ch:readoutnoise}

One displacement noise contribution---and by design often the limiting one---is noise in the heterodyne signal read out, particularly noise in the electric current of the photo detector that measures the interference signal of received and local laser beams.
The carrier-to-noise-density ratio $C/N_0$ (in units of power spectral density) can be used to calculate the resulting phase noise $\widetilde{\phi}_\text{r/o}$ in units of rad/$\sqrt{\text{Hz}}$ (linear spectral density):
\begin{equation}
\widetilde{\phi}_\text{r/o} \left[ \frac{\text{rad}}{\sqrt{\text{Hz}}} \right] = \frac{1}{\sqrt{C/N_0}} ~ .
\label{eq:oneoversnr}
\end{equation}

$C$ corresponds to the signal power, and the amplitude $\sqrt{C}$ can be expressed as electric current
\begin{equation}
I_\text{total} = R_\text{pd}\:\frac{P_\text{total}}{2N_\text{pd}} ~ ,
\label{eq:totalincidentcurrent}
\end{equation}
which is proportional to the time-dependent total incident optical power
\begin{equation}
\begin{split}
P_\text{total} = ~ & \overbrace{P_\text{local} + P_\text{rec}}^{\text{DC term}} ~ + \\
&  \underbrace{\overbrace{2\sqrt{\eta_\text{het} P_\text{local} P_\text{rec} }}^{\text{amplitude}} \:\overbrace{\sin\left(2\pi f_\text{het}t+\varphi\right)}^{\text{time dependence}}}_\text{\text{AC term (heterodyne beat note)}}
\end{split}
\label{eq:totalincidentpower}
\end{equation}
where $P_\text{local}$ is the power of the local laser. Dropping the DC term and the time dependence, the RMS electrical signal for the heterodyne beat note on one segment of a photodiode is found as
\begin{equation}
I_\text{signal, rms} = \frac{1}{\sqrt{2}}\:R_\text{pd}\frac{2\sqrt{\eta_\text{het} P_\text{local} P_\text{rec} }}{2N_\text{pd}} ~ .
\label{eq:totalincidentrms}
\end{equation}
The factor $2N_\text{pd}$ accounts for the fact that there are two output ports of the 50:50 beam splitter that combines the received laser light with the local laser, and each beam is distributed over $N_\text{pd}$ segments of the photodiode.\\

$N_0$ corresponds to the power spectral density, and the single-sided linear spectral density $\sqrt{N_0}$ can be expressed as the electric current noise $\widetilde{I}$ in units of A$/\sqrt{\text{Hz}}$. It is composed of
\begin{enumerate} 
	\item \textbf{shot noise}, the fluctuations of the number of photons detected,
	\item \textbf{relative intensity noise}, the fluctuations in the laser power, and
	\item \textbf{electrical noise}, the residual noise introduced by the transimpedance amplifier.
\end{enumerate}
We will now determine the individual noise contribution for each component.

\subsubsection{Shot Noise}

For our purpose it is sufficient to compute the shot noise based on the DC term found in Equation~(\ref{eq:totalincidentpower}) which leads to a total average DC photocurrent
\begin{equation}
I_\text{dc} \approx R_\text{pd}\frac{P_\text{local}+P_\text{rec}}{2N_\text{pd}} ~.
\label{eq:totalincidentdc}
\end{equation}
With $q_\text{e}$ as the electron charge the shot noise can now be expressed as
\begin{equation}
\widetilde{I}_\text{sn} = \sqrt{2\: q_\text{e}\: I_\text{dc}} \approx \sqrt{2\:q_\text{e}\:R_\text{pd}\frac{P_\text{local}+P_\text{rec}}{2N_\text{pd}}}
\label{eq:shotnoise}
\end{equation}
with minor corrections to be found in \cite{meers1991modulation, niebauer1991nonstationary}.
Following Equation~(\ref{eq:oneoversnr}) the read-out noise due to shot noise is
\begin{equation}
\widetilde{\phi}_\text{r/o}^\text{sn}  = \frac{\widetilde{I}_\text{sn}}{I_\text{signal}} = \sqrt{\frac{2N_\text{pd} \: q_\text{e} \left(P_\text{local}+P_\text{rec}\right)}{R_\text{pd}\: \eta_\text{het}\: P_\text{local}\: P_\text{rec}}}
\label{eq:readoutshotnoise}
\end{equation} 
for each photodiode segment. For sufficient high values of $P_\text{local}/P_\text{rec}$, a higher total local laser power may have no influence on the shot noise in the signal read-out. 

Shot noise is a non-correlated contribution between different photodiodes and segments, hence averaging over all $N_\text{pd}$ segments will improve the signal quality by a factor of $\sqrt{N_\text{pd}}$ so it becomes independent of the number of segments (single-element vs. quadrant photodiode):
\begin{equation}
\left<\widetilde{\phi}_\text{r/o}^\text{sn}\right>  = \frac{1}{\sqrt{N_\text{pd}}}\: \widetilde{\phi}_\text{r/o}^\text{sn} = \sqrt{\frac{2 \: q_\text{e} \left(P_\text{local}+P_\text{rec}\right)}{R_\text{pd}\: \eta_\text{het}\: P_\text{local}\: P_\text{rec}}} ~ .
\label{eq:meanreadoutshotnoise}
\end{equation}

\subsubsection{Relative Intensity Noise}

The relative intensity noise, $RIN$, as described in Section~\ref{sec:optelec} is uncorrelated between both laser beams. It couples directly to the photocurrent and adds quadratically:
\begin{equation}
\begin{split}
\widetilde{I}_\text{rin} ~ & =  \sqrt{\left(R_\text{pd} \frac{P_\text{local}}{2N_\text{pd}} RIN \right)^2 + \left(R_\text{pd} \frac{P_\text{rec}}{2N_\text{pd}}RIN\right)^2}\\
& = R_\text{pd}\frac{\sqrt{P_\text{local}^2 + P_\text{rec}^2}}{2N_\text{pd}}\: RIN~.
\label{eq:rinnoise}
\end{split}
\end{equation}
Consequently, the read-out noise due to relative intensity noise is
\begin{equation}
\widetilde{\phi}_\text{r/o}^\text{rin}  = \frac{\widetilde{I}_\text{rin}}{I_\text{signal}} = RIN\:\sqrt{\frac{P_\text{local}^2+P_\text{rec}^2}{2\eta_\text{het}\:P_\text{local}\:P_\text{rec}}}
\label{eq:readoutrinnoise}
\end{equation} 
for each photodiode segment, generally independent of the number of segments and the photodiode responsivity. A higher local laser power may increase the influence of relative intensity noise in the signal read-out.

Since the relative intensity noise is correlated in both beam splitter outputs and on each photodiode segment, averaging over photodiodes or $N_\text{pd}$ segments does not yield any improvements in the signal quality:
\begin{equation}
\left<\widetilde{\phi}_\text{r/o}^\text{rin}\right>  = \widetilde{\phi}_\text{r/o}^\text{rin} =  RIN\:\sqrt{\frac{P_\text{local}^2+P_\text{rec}^2}{2\eta_\text{het}\:P_\text{local}\:P_\text{rec}}} ~ .
\label{eq:meanreadoutrinnoise}
\end{equation}

\subsubsection{Electrical Noise}

The photodiode preamplifier (transimpedance amplifier) shows input current noise, $\widetilde{I}_\text{pd}$, as well as uncorrelated voltage noise, $\widetilde{U}_\text{pd}$, that can be converted to equivalent input current noise $\widetilde{I}_\text{tia} = \widetilde{U}_\text{pd}/Z_\text{pd}$ using the photodiode's impedance $Z_\text{pd}$. Both contributions add quadratically.
\begin{equation}
\widetilde{I}_\text{el}  =  \sqrt{\widetilde{I}_\text{pd}^2 + \widetilde{I}_\text{tia}^2}
= \sqrt{\widetilde{I}_\text{pd}^2 + \left(\frac{\widetilde{U}_\text{pd}}{Z_\text{pd}}\right)^2}
\label{eq:electronicnoise}
\end{equation}
The read-out noise due to electronic noise is then given by
\begin{equation}
\widetilde{\phi}_\text{r/o}^\text{el} = \frac{\widetilde{I}_\text{el}}{I_\text{signal}} = N_\text{pd}\:\frac{\sqrt{2}}{R_\text{pd}}\:\sqrt{\frac{\widetilde{I}_\text{pd}^2 + \left(\frac{\widetilde{U}_\text{pd}}{Z_\text{pd}}\right)^2}{\eta_\text{het}\:P_\text{local}\:P_\text{rec}}}
\label{eq:readoutelectronicnoise}
\end{equation} 
for each photodiode segment. Here, a higher local laser power will reduce the influence of electronic noise in the signal read-out.

Electronic noise is a non-correlated contribution between different photodiodes and segments, hence averaging over all $N_\text{pd}$ segments will improve the signal quality by a factor of $\sqrt{N_\text{pd}}$. As a result, the influence of electronic noise in the signal read-out scales by $\sqrt{N_\text{pd}}$ since each channel is amplified individually:
\begin{equation}
\left<\widetilde{\phi}_\text{r/o}^\text{el}\right>  = \frac{1}{\sqrt{N_\text{pd}}}\: \widetilde{\phi}_\text{r/o}^\text{el} = \frac{\sqrt{2N_\text{pd}}}{R_\text{pd}}\:\sqrt{\frac{\widetilde{I}_\text{pd}^2 + \left(\frac{\widetilde{U}_\text{pd}}{Z_\text{pd}}\right)^2}{\eta_\text{het}\:P_\text{local}\:P_\text{rec}}} ~ .
\label{eq:meanreadoutelectronicnoise}
\end{equation}

\subsubsection{Optimal Local Laser Power}\label{sec:optimallaser}
\begin{figure}[htb!]\centering
	\includegraphics[width=.675\linewidth]{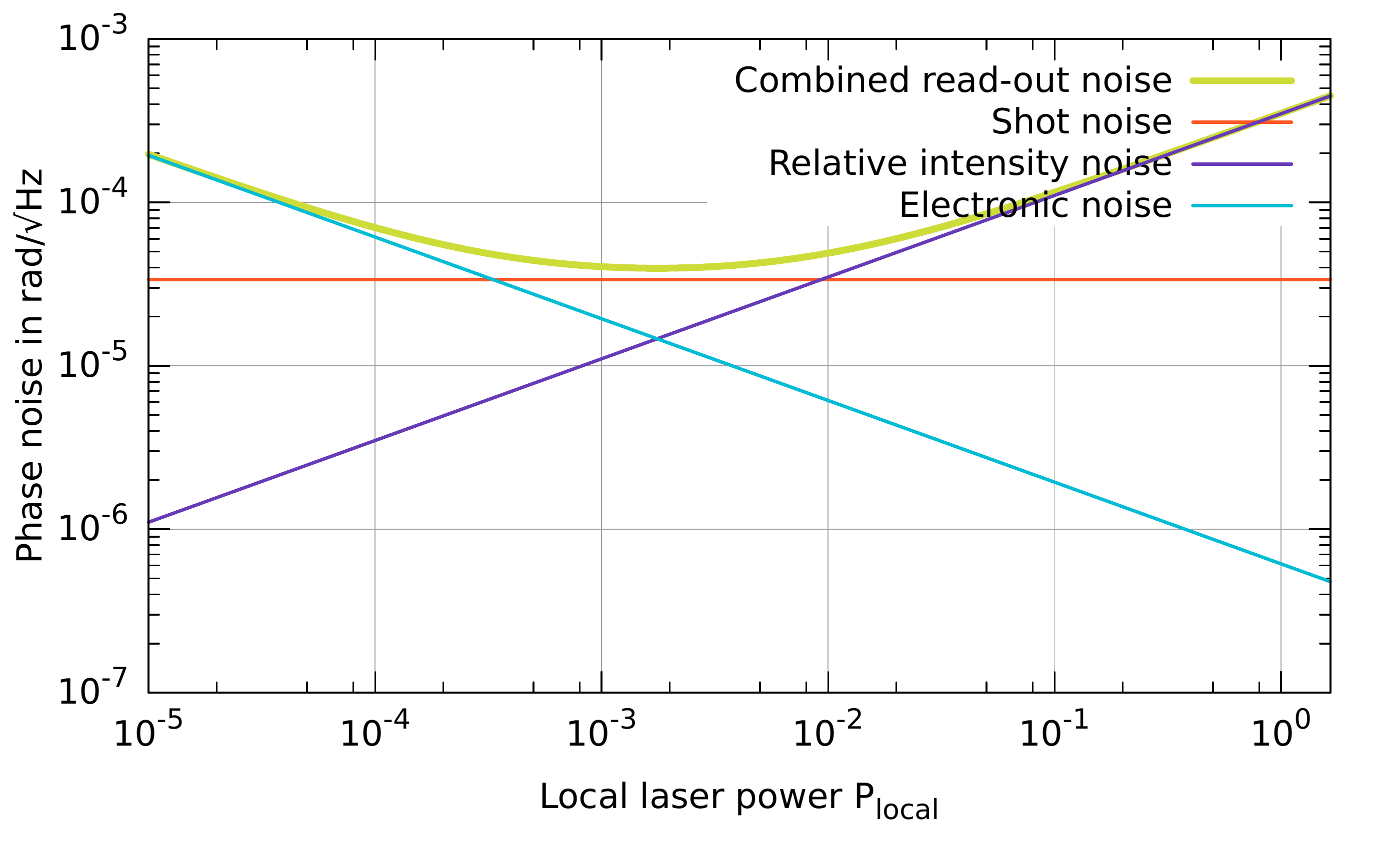}
	\caption{Linear spectral density of combined read-out phase noise (green) and its individual contributions over local laser power $P_\text{local}$.}
	\label{fig:readout-noise}
\end{figure}

As mentioned above, the influence of the different read-out noise contributions scales differently with local laser power $P_\text{local}$. Figure~\ref{fig:readout-noise} shows the total read-out noise
\begin{equation}
\left<\widetilde{\phi}_\text{r/o}^\text{total}\right> = \sqrt{
	\left<\widetilde{\phi}_\text{r/o}^\text{sn}\right>^2 +
	\left<\widetilde{\phi}_\text{r/o}^\text{rin}\right>^2 +
	\left<\widetilde{\phi}_\text{r/o}^\text{el}\right>^2}
\label{eq:totalmeanreadoutnoise}
\end{equation}
as well as the individual contributions for the given parameters plotted over local laser power. A minimum of this function can be found for $P_\text{local} = \num{1.75e-3}$\,Watts.\\

\begin{figure}[htb!]\centering
	\includegraphics[width=.675\linewidth]{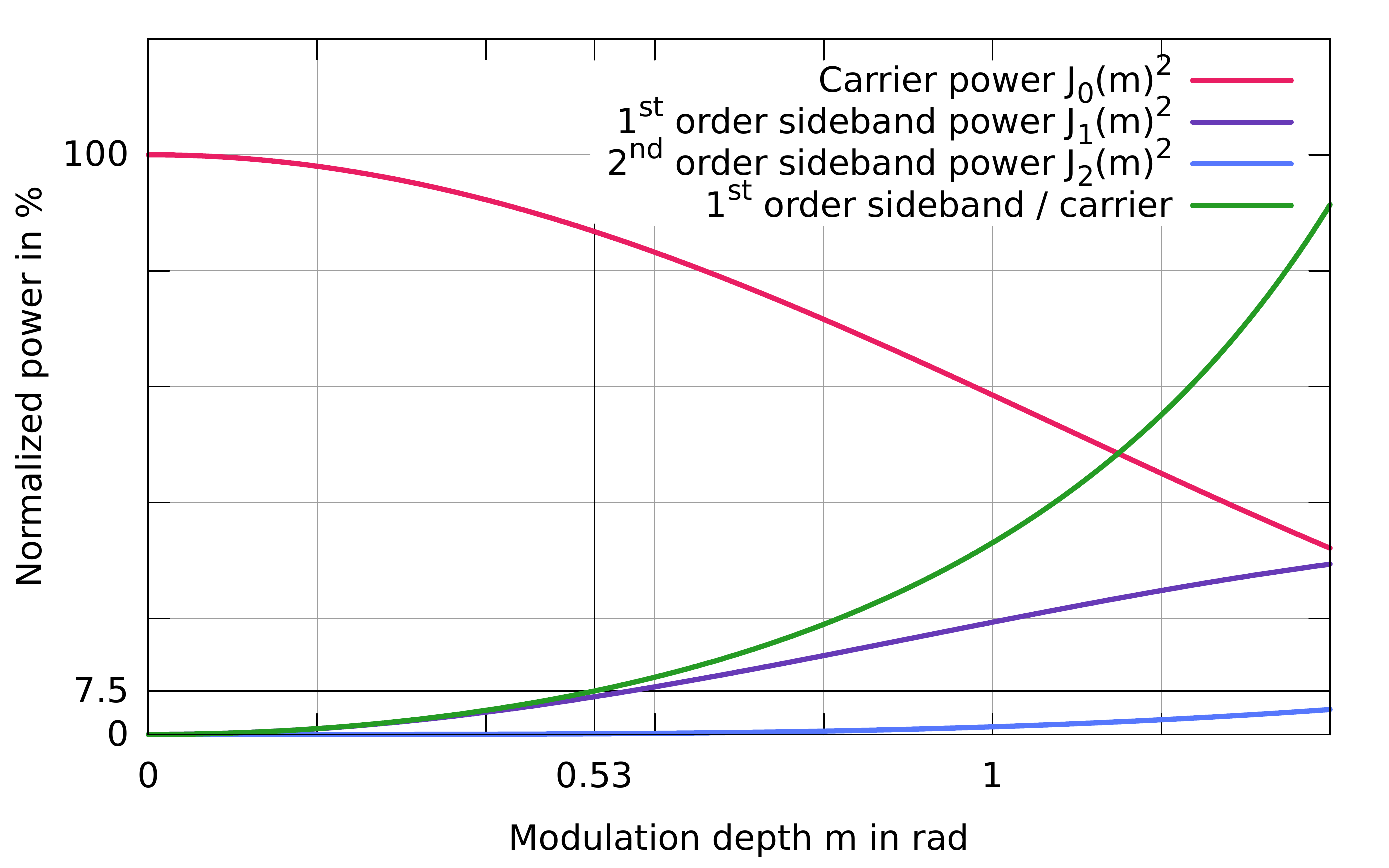}
	\caption{Carrier, first- and second-order sidebands (normalized power over modulation depth $m$). The desired ratio between carrier and first-order sideband (green trace) of 7.5\% occurs at $m=0.53$\,rad as indicated.}
	\label{fig:modulationdepth}
\end{figure}

Before we can compute the absolute values for the different read-out noise contributions, we have to consider that in reality the laser beams are phase modulated and carry additional information in sidebands. As a result, the heterodyne signal now consists of a carrier beat note and multiple sideband beat-notes. These sidebands consume some of the total signal power. In the present case we require each of the two first-order sidebands to hold 7.5\% of the carrier's power.\footnote{Additional signal modulation used for inter-spacecraft data transfer and ranging is assumed to contain approximately \num{1}\% of the signal power (see \cite{esteban2011experimental}) and thus can be ignored at this point.} The resulting frequency spectrum can be calculated using Bessel functions of the first kind (${J}_{0}$, ${J}_\text{1}$, ${J}_\text{2}$, ...). Figure~\ref{fig:modulationdepth} shows the power for the carrier (${J}_{0}(m)^2$) and the first- and second-order sidebands (${J}_\text{1}(m)^2$, ${J}_\text{2}(m)^2$) as fractions of the total power as a function of the modulation depth $m$. The desired ratio between carrier and first-order sideband of 7.5\% occurs at $m=0.53$\,rad.\footnote{High-power first-order sidebands that result in a modulation depth $m > 1$ will additionally be accompanied by higher order sidebands. This should be avoided since these sidebands are not used but nevertheless reduce the overall signal power.}

Accordingly the RMS electrical signal for the carrier beat note has to be written as 
\begin{equation}
\begin{split}
I_\text{carrier} ~& = \frac{1}{\sqrt{2}}\:R_\text{pd}\frac{2\sqrt{\eta_\text{het} \: {J}_{0}(m)^2\: P_\text{local} \: {J}_{0}(m)^2\: P_\text{rec} }}{2N_\text{pd}}\\
& = {J}_{0}(m)^2 \: I_\text{signal}
\label{eq:carrierincidentrms}
\end{split}
\end{equation}
and we must apply this reduced carrier signal level to the read-out noise calculations. As obvious from Equations~(\ref{eq:readoutshotnoise}), (\ref{eq:readoutrinnoise}) and (\ref{eq:readoutelectronicnoise}), the individual noise contributions are simply increased by the factor $1/{J}_{0}(m)^2 = \num{1.15}$. Converted to displacement noise (see Equation~(\ref{eq:radinpm})) we finally obtain 
\begin{equation}
\begin{split}
\left<\widetilde{x}_\text{r/o}^\text{sn}\right>_\text{carrier} ~& = \frac{\lambda_\text{laser}}{2\pi}\:\frac{1}{{J}_{0}(m)^2}\: \left<\widetilde{\phi}_\text{r/o}^\text{sn}\right>\\
& = \num{6.58e-12} \frac{\text{m}}{\sqrt{\text{Hz}}} ~,
\end{split}
\label{eq:meancarrierreadoutshotnoise}
\end{equation}
\begin{equation}
\begin{split}
\left<\widetilde{x}_\text{r/o}^\text{rin}\right>_\text{carrier} ~& = \frac{\lambda_\text{laser}}{2\pi}\:\frac{1}{{J}_{0}(m)^2}\: \left<\widetilde{\phi}_\text{r/o}^\text{rin}\right>\\
& = \num{2.85e-12} \frac{\text{m}}{\sqrt{\text{Hz}}} ~\text{, and}
\end{split}
\label{eq:meancarrierreadoutrinnoise}
\end{equation}
\begin{equation}
\begin{split}
\left<\widetilde{x}_\text{r/o}^\text{el}\right>_\text{carrier} ~& = \frac{\lambda_\text{laser}}{2\pi}\:\frac{1}{{J}_{0}(m)^2}\: \left<\widetilde{\phi}_\text{r/o}^\text{el}\right>\\
& = \num{2.86e-12} \frac{\text{m}}{\sqrt{\text{Hz}}}~.
\end{split}
\label{eq:meancarrierreadoutelectronicnoise}
\end{equation}

From the values above (and also clearly visible in Figure~\ref{fig:readout-noise}) we conclude that the total read-out noise in the carrier signal,
\begin{equation}
\begin{split}
\left<\widetilde{x}_\text{r/o}^\text{total}\right>_\text{carrier} ~& = \frac{\lambda_\text{laser}}{2\pi}\:\frac{1}{{J}_{0}(m)^2}\: \left<\widetilde{\phi}_\text{r/o}^\text{total}\right>\\
& = \num{7.72e-12} \frac{\text{m}}{\sqrt{\text{Hz}}} ~,
\end{split}
\label{eq:meancarrierreadouttotalnoise}
\end{equation}
is limited by shot noise as desired by a carefully designed gravitational wave observatory. This value is equivalent to a phase noise of \num{4.56e-05}\,$\text{rad}/{\sqrt{\text{Hz}}}$. One usually aims to keep additional phase fluctuations of the signal as well as all noise introduced during phase measurement, post-processing and data analysis well below this level.

\subsection{Clock Noise}

To measure the phase of the carrier signal, the analog output from the transimpedance amplifier is digitized by an analog-to-digital converter (ADC) that is triggered by a reference oscillator (system clock). Here, timing noise $\widetilde{t}$ leads to phase noise $\widetilde{\phi} = 2\pi\:f\:\widetilde{t}$ in the digital representation of the signal. For the measurement of a signal with frequency $f=f_\text{het}$ a timing stability of $\widetilde{t} < \num{4.03e-13}\,\text{s}/{\sqrt{\text{Hz}}}$ would be required to stay below the above calculated total carrier signal read-out phase noise. Unfortunately, ADCs and oscillators that stable do not exist. To deal with the excess noise, additional signals called `pilot tones' are introduced. Within one spacecraft, a common pilot tone (at frequency $f_p$ outside the heterodyne signal bandwidth) is combined with the analog output from each transimpedance amplifier. Both signals are digitized simultaneously in each ADC channel. Thus we can use the pilot tone as a reference to suppress the influence of ADC timing jitter on the digitized carrier signal.

\begin{figure}[htb!]\centering 
	\includegraphics[width=\linewidth]{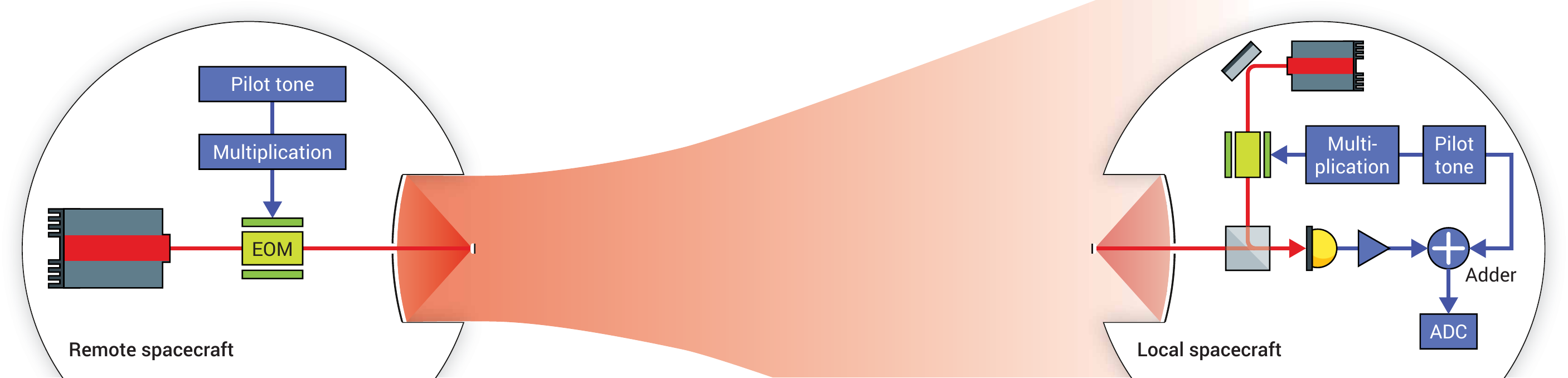}
	\caption{Pilot tone distribution for a single link of the observatory. At the remote spacecraft, the pilot tone frequency is multiplied and the signal is modulated onto the outgoing laser beam by an electro-optic phase modulator (EOM). A separate pilot tone on the local spacecraft is modulated onto the local laser to compare the two pilot tones in the sideband beat note of the heterodyne signal. To suppress the influence of ADC timing jitter, the local pilot tone is added to the heterodyne signal and used as a reference.}
	\label{fig:cttc}
\end{figure}

In addition, the pilot tones of different spacecraft are modulated on the outgoing laser beams by electro-optic phase modulators (EOMs) as illustrated in Figure~\ref{fig:cttc}. The affiliated first-oder sidebands (as already mentioned in Section~\ref{sec:optimallaser}) each hold 7.5\% of the carrier's power and result in sideband beat notes in the heterodyne signal. These additional beat notes (which must fall within the heterodyne signal bandwidth) are correlated with the differential phase noise between the corresponding remote and local pilot tones. Thus we can compare the pilot tones between all spacecraft and construct a constellation wide common reference during post-processing. As a result, a specific timing stability of the individual system clocks is no longer required, but the technique is limited by
\begin{enumerate} 
	\item \textbf{read-out noise} in the sideband beat notes, and 
	\item excess phase noise introduced by components in the \textbf{pilot tone transmission chain}.
\end{enumerate}
We will now calculate the corresponding displacement noise contributions for both.

\subsubsection{Sideband Signal Read-out Noise}

Since the RMS electrical signal for the sideband beat note
\begin{equation}
I_\text{sideband} = 7.5\% \times I_\text{carrier} = J_1(m)^2\:I_\text{signal}
\label{eq:sidebandincidentrms}
\end{equation}
is smaller than the carrier signal (compare Equation~(\ref{eq:carrierincidentrms})), the read-out phase noise for the sideband signal will be much higher (compare Equation~(\ref{eq:meancarrierreadouttotalnoise})). To reduce the impact of read-out noise on the sideband signals, we boost the desired signal---which is the pilot tone's phase information---before modulating it onto the laser beams. This can be done by frequency multipliers as they conserve timing jitter\footnote{The timing jitter conservation of frequency multipliers and dividers stands in contrast to the mixing process in, e.g., heterodyne interferometry or electronic mixers, which maintains phase information.} and hence lead to an amplification of phase jitter by the frequency multiplication (signal amplification) factor $f_\text{mod}/f_p$ where $f_\text{mod}$ represents the actual modulation frequency.

Accordingly the total read-out noise for one first-order sideband beat note expressed in phase noise,
\begin{equation}
\left<\widetilde{\phi}_\text{r/o}^\text{total}\right>_\text{sideband} = \frac{f_\text{p}}{f_\text{mod}} \: \frac{1}{{J}_\text{1}(m)^2}\: \left<\widetilde{\phi}_\text{r/o}^\text{total}\right> ~,
\label{eq:meansidebandsreadouttotalnoisephase}
\end{equation}
scales with the inverse of the signal amplification factor. Furthermore, the equivalent displacement noise after Equation~(\ref{eq:radinpm}) scales with the ratio of the maximum heterodyne frequency $f_\text{het}$ to the pilot tone frequency $f_p$ as
\begin{equation}
\left<\widetilde{x}_\text{r/o}^\text{total}\right>_\text{sideband} = \frac{\lambda_\text{laser}}{2\pi} \: \frac{f_\text{het}}{f_p} \left<\widetilde{\phi}_\text{r/o}^\text{total}\right>_\text{sideband}
\label{eq:meansidebandsreadouttotalnoisepm}
\end{equation}
since all measurements (at frequency $f_\text{het}$) are referenced to the pilot tone. The higher the pilot tone frequency, the less phase jitter of the pilot tone impacts the measurement of a signal, and the higher the signal frequency, the more it is influenced by phase jitter of the pilot tone.

In conclusion, the total read-out noise for both sidebands combined (factor $1/\sqrt{2}$) is
\begin{equation}
\begin{split}
\left<\widetilde{x}_\text{r/o}^\text{total}\right>_\text{sidebands} ~& = \frac{1}{\sqrt{2}}\:\frac{\lambda_\text{laser}}{2\pi}\:\frac{f_\text{het}}{f_\text{mod}}\:\frac{1}{{J}_\text{1}(m)^2}\: \left<\widetilde{\phi}_\text{r/o}^\text{total}\right>\\
& = \num{5.46e-13}\,\frac{\text{m}}{\sqrt{\text{Hz}}}
\end{split}
\label{eq:meansidebandsreadouttotalnoise}
\end{equation}
for a modulation frequency of $f_\text{mod} = \num{2.4}$\,GHz. This value represents the excess noise of the observatory introduced by the imperfect synchronization of the pilot tones between the different spacecraft due to the noisy read-out of the sideband signal. The pilot tone frequency $f_p$ does not influence this  noise level but might be of importance during the actual phase measurement and in the generation of the modulation signal.

\subsubsection{Pilot Tone Transmission Chain Noise}

Another source of excess noise is induced by a reduced pilot tone fidelity, that is when the phase of the modulation sidebands differs from the phase of the corresponding pilot tones used for the ADC timing jitter correction. Components in the pilot tone transmission chain might shift the phase of the pilot tone (in the electrical signal) or sidebands (in the optical signal). There are many components involved that can potentially limit the observatories sensitivity in this way.\\

\begin{figure}[htb!]\centering
	\includegraphics[width=.675\linewidth]{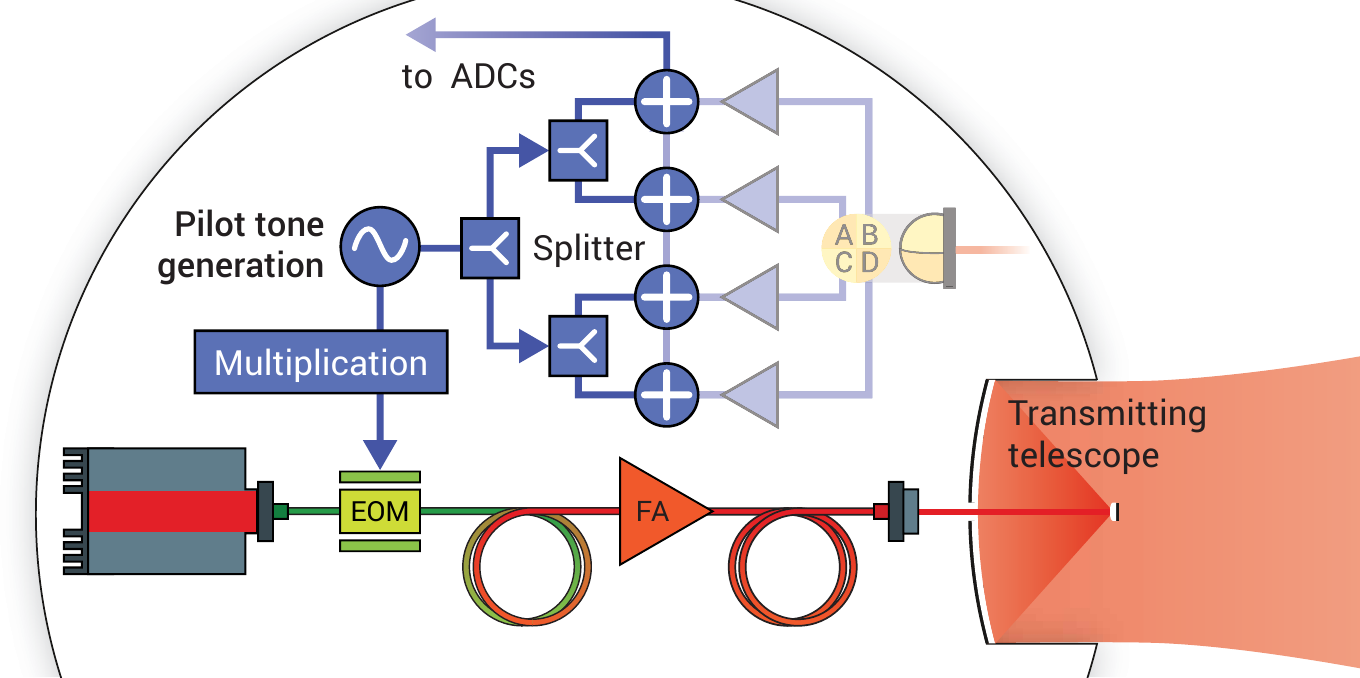}
	\caption{The pilot tone (that is combined with the heterodyne signals and used as a reference to suppress timing jitter) must be phase stable to the sidebands that are modulated onto the outgoing laser beam by an EOM. Components in the transmission line from the pilot tone generation to the ADCs (like power splitters and adders) and to the transmitting telescope (like fiber amplifiers (FA) and optical fibers) might add phase noise.}
	\label{fig:line}
\end{figure}

As illustrated in Figure~\ref{fig:line} (blue items) the electrical transmission chain contains a number of power splitters, the power combiner (adder) for the pilot tone and the heterodyne signal, as well as the frequency multiplier (or divider), possibly in multiple stages. Since phase noise introduced by any of these components depends on the actual pilot tone frequency, the combined noise introduced by all electrical components in the pilot tone transmission chain, 
$\widetilde{t}_\text{el} = \num{4e-14}\,\text{s}/{\sqrt{\text{Hz}}}$,
is given in frequency independent units of timing jitter. This translates to an equivalent displacement noise of
\begin{equation}
\widetilde{x}_\text{tml}^\text{el} = \lambda_\text{laser}\:f_\text{het}\:\widetilde{t}_\text{el} = \num{7.66e-13}\,\frac{\text{m}}{\sqrt{\text{Hz}}} ~.
\label{eq:pttimingjitterx}
\end{equation}
Keep in mind that the above values, like most noise figures given in this document, depend on the temperature stability. Actual dependencies for individual components may change with temperature and can also (partly) cancel each other. Thus a complete timing noise model for all electrical components would turn out to be quite complex.\\

Also electrical cables connecting the different components shift the phase of the pilot tone and modulation signal in accordance with temperature due to a number of effects, among others a change in the dielectric constant of the inner insulator and a change in the cables' dimension. This will alter the velocity of propagation and the electrical length of the transmission line respectively.

The absolute phase shift depends on the actual frequency of the signal passed along the cable, and different frequencies ($f_p$, $f_\text{mod}$) are involved. However, with a thermal stability of the electrical cables given per meter and gigahertz, we can calculate an equivalent displacement noise level independent of the signal frequency. This thermal stability is assumed to be 
\begin{equation}
\left(\frac{\delta \phi}{\delta T}\right)_\text{cables} =  \num{7}\,\frac{\text{mrad}}{\text{K}}\:\frac{1}{\text{m} \times \text{GHz}}
\label{eq:thermalstabilitycables}
\end{equation}
and leads to a noise due to temperature shifts in the electrical cables (given by the temperature noise at the electronics and electro-optics $\widetilde{T}_\text{el}\left(f\right)$, compare Section~\ref{sec:temperature}) of
\begin{equation}
\begin{split}
\widetilde{x}_\text{tml}^\text{cables}\left(f\right) ~&= \frac{\lambda_\text{laser}}{2\pi}\:f_\text{het}\:\widetilde{T}_\text{el}\left(f\right)\:l_\text{cables}\left(\frac{\delta \phi}{\delta T}\right)_\text{cables} \\
&= \num{4.27e-11}\,\frac{\text{m}}{\text{K}} \times \widetilde{T}_\text{el}\left(f\right) 
\label{eq:cablespm}
\end{split}
\end{equation}
that changes with Fourier frequency $f$. The length of the electrical cables was assumed to be $l_\text{cables} = \num{2}$\,m. In the above equation, the signal frequency ($f_p$, $f_\text{mod}$) canceled with parts of the corresponding scaling factor introduced by Equation~(\ref{eq:meansidebandsreadouttotalnoisepm}), and only the maximum heterodyne frequency $f_\text{het}$ remains.\\

Likewise, the influence of optical fibers that pass the modulated laser light from the EOM to the transmitting telescope (see Figure~\ref{fig:line}) can be calculated. Here, the modulation signal is phase shifted with temperature due to a change in the fibers' dimension and refractive index. For a thermal stability of the fibers given per meter and gigahertz,
\begin{equation}
\left(\frac{\delta \phi}{\delta T}\right)_\text{fibers} =  \num{1}\,\frac{\text{mrad}}{\text{K}}\:\frac{1}{\text{m} \times \text{GHz}} ~,
\label{eq:thermalstabilityfibers}
\end{equation}
and a total fiber length, $l_\text{fibers} = \num{5}$\,m, the equivalent displacement noise due to temperature shifts in the optical fibers (given by the same temperature noise at the electronics and electro-optics $\widetilde{T}_\text{el}\left(f\right)$, compare Section~\ref{sec:temperature}) is
\begin{equation}
\begin{split}
\widetilde{x}_\text{tml}^\text{fibers}\left(f\right) ~&= \frac{\lambda_\text{laser}}{2\pi}\:f_\text{het}\:\widetilde{T}_\text{el}\left(f\right)\:l_\text{fibers}\left(\frac{\delta \phi}{\delta T}\right)_\text{fibers} \\
&= \num{1.52e-11}\,\frac{\text{m}}{\text{K}} \times \widetilde{T}_\text{el}\left(f\right) ~.
\label{eq:fiberspm}
\end{split}
\end{equation}

Finally, the two electro-optic components that sit in the optical transmission chain, namely the EOM and a fiber amplifier (FA) that boosts the laser power to $> P_\text{tel}$ before passing it to the telescope, can influence the phase of the sidebands. The performance of both devices depends on the absolute temperature, light power, temperature stability and other environmental influences and should be subject to a separate study. We assumed a phase noise of $\widetilde{\phi}_\text{eom}=\num{0.0003}\,\text{rad}/{\sqrt{\text{Hz}}}$ for the EOM and $\widetilde{\phi}_\text{fa}=\num{0.0006}\,\text{rad}/{\sqrt{\text{Hz}}}$ for the FA, valid at the modulation frequency $f_\text{mod}$.\\

Since all noise sources are temperature dependent, we conservatively add the individual figures linearly and come up with a total pilot tone transmission chain noise of
\begin{equation}
\begin{split}
\widetilde{x}_\text{tml}^\text{total}\left(f\right) = ~& \widetilde{x}_\text{tml}^\text{el} + \widetilde{x}_\text{tml}^\text{cables}\left(f\right) + \widetilde{x}_\text{tml}^\text{fibers}\left(f\right) \\
& + \frac{\lambda
	_\text{laser}}{2\pi} \left(\widetilde{\phi}_\text{eom} + \widetilde{\phi}_\text{fa}\right).\\
\label{eq:transmissionlinepm}
\end{split}
\end{equation}

\begin{figure}[htb!]\centering
	\includegraphics[width=.675\linewidth]{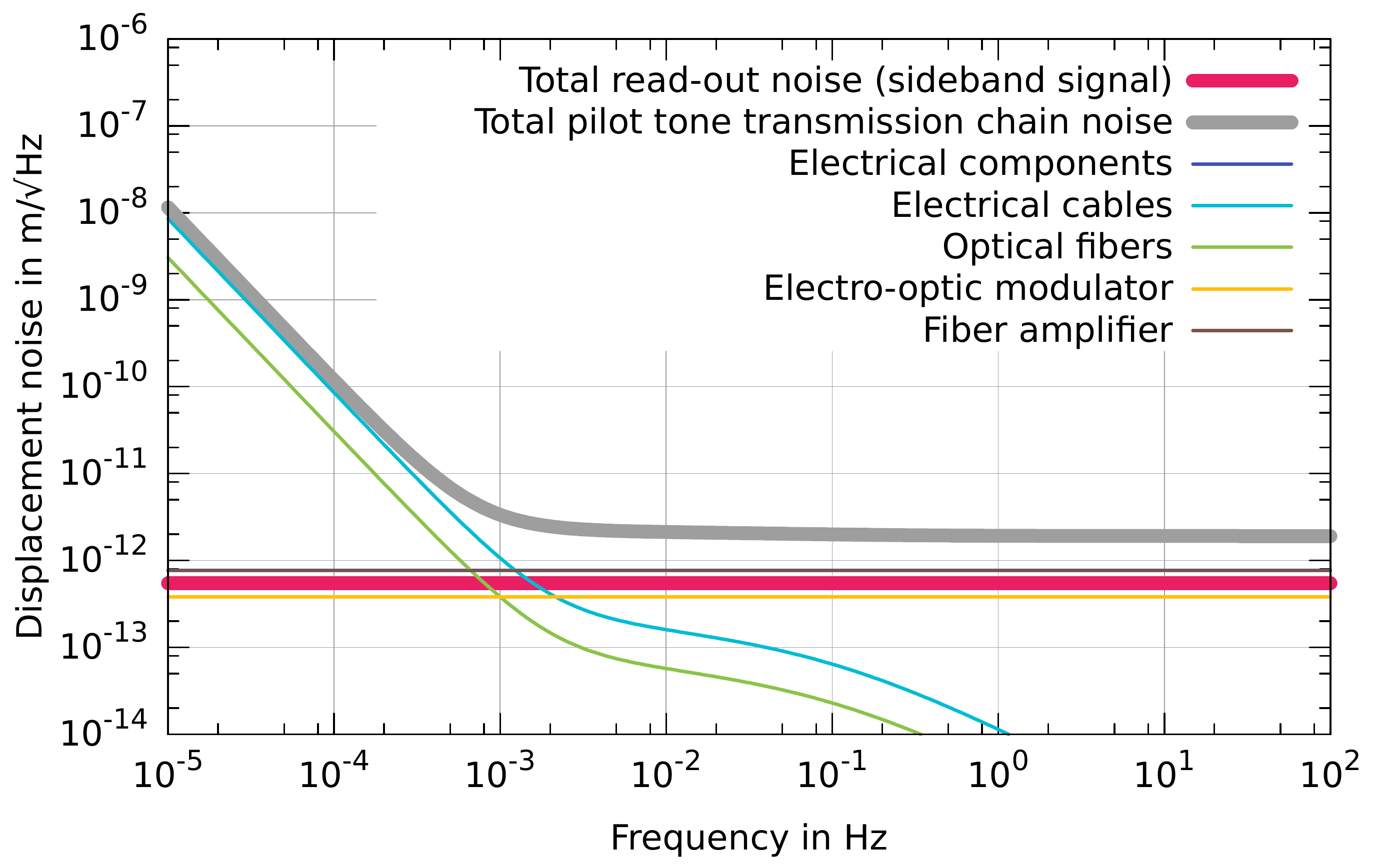}
	\caption{Clock noise contributions plotted over Fourier frequency, including the sideband read-out noise (for both sideband signals combined) and the individual pilot tone transmission chain noise components.}
	\label{fig:clock-noise}
\end{figure}

Figure~\ref{fig:clock-noise} shows all noise contributions individually over Fourier frequency $f$. At low frequencies, the importance of the consideration of temperature noise becomes obvious since it clearly dominates the equivalent displacement noise.

\subsection{Optical Path Length Noise}\label{ch:pathlengthnoise}

The optical telescopes are naturally within the optical path of the interferometer and jitter of the telescope length directly translates to optical path length noise. The dimensional jitter is caused by temperature noise at the telescope, but it is hard to model because of a strong temperature gradient. While the primary mirror usually lies deep within the spacecraft and could be close to room temperature, the secondary mirror is more exposed to outer space and may be as cold as a few Kelvin. Dimensional stability investigations for carbon fiber reinforced polymer and ultra-low expansion glass-ceramic structures reached a path length noise smaller than $\widetilde{x}_\text{opn}^{tel} = \num{1}$\,$\text{pm}/{\sqrt{\text{Hz}}}$ down to frequencies of 1\,mHz \cite{preston2010stability}.\\

Also a change in temperature of the optical bench (fused silica components bonded to a base plate made out of base plate made out of a thermally-compensated glass-ceramic) results in a uniform expansion of the material that leads to a phase shift in the heterodyne signals. If  more than one interferometer is located on a single optical bench, this effect will only cancel out if the path length on the optical bench is the same for all interferometers. If there is a path length imbalance, however, the  phase noise due to temperature fluctuations will not cancel completely. Instead, there will be a coupling factor that scales with the difference in the optical path lengths of at least two interferometers involved.

As discussed in Section~\ref{sec:opticalbench}, a dedicated inter-spacecraft interferometer is needed to utilize the full power of the received beam and minimize the influence of read-out noise. In this read-out scheme, the  influence of the optical path length difference in the combination of any two inter-spacecraft interferometers (for one full observatory arm) cancels each other. Accordingly the relevant path length difference is the one between the two additional interferometers required to determine the proof mass displacement: the  proof mass interferometer and the reference interferometer, compare Figure~\ref{fig:opd}.

One must distinguish between the path length difference within fused silica, $\text{OPD}_\text{fs}$, and the path length difference on the optical bench itself, $\text{OPD}_\text{ob}$. We assume values of $\text{OPD}_\text{fs} = \num{29}$\,mm and $\text{OPD}_\text{ob} = \num{565}$\,mm. The latter is the total path length difference on the optical bench including light paths within fused silica optics, so the significant path length difference on the glass-ceramic base plate comes down to $\text{OPD}_\text{ob} - \text{OPD}_\text{fs}$.
We can now calculate the equivalent displacement noise contributions due to the path length imbalances. With the given temperature noise at the optical bench, $\widetilde{T}_\text{ob}\left(f\right)$, and the coefficient of thermal expansion of glass-ceramic, $\alpha_\text{ule} = \num{2.0e-08}$\,$\text{m}/\text{K}$, the path length noise of the base plate can be expressed as
\begin{equation}
\widetilde{x}_\text{opn}^\text{ule}\left(f\right) = \widetilde{T}_\text{ob}\left(f\right)\times\left(\text{OPD}_\text{ob} - \text{OPD}_\text{fs}\right)\times\alpha_\text{ule}
\label{eq:xule}
\end{equation}

The description of the path length noise introduced by the fused silica components is more complex since the laser beam is passing through those components and not through vacuum. Thus we have to consider the refractive index of fused silica, $n_\text{fs} = 1.45$, as well as its change with temperature, $dn_\text{fs}/dT = \num{1.1e-6}/\text{K}$.\footnote{The given values are only valid for a wavelength of 1064\,nm.} With the coefficient of thermal expansion of fused silica, $\alpha_\text{fs} = \num{5.5e-7}$\,$\text{m}/\text{K}$, the equivalent displacement noise can then be expressed as
\begin{equation}
\widetilde{x}_\text{opn}^\text{fs}\left(f\right) = \widetilde{T}_\text{ob}\left(f\right)\times\text{OPD}_\text{fs} \left(\alpha_\text{fs}\left(n_\text{fs}-1\right)+\frac{dn_\text{fs}}{dT}\right)
\label{eq:xfs}
\end{equation}
In the above equation we use the difference of the refractive index of fused silica and vacuum, $n_\text{fs}-1$. This is due to the fact that an increase in the path length for light passed through fused silica simultaneously decreases the path length in vacuum.\\

\begin{figure}[htb!]\centering
	\includegraphics[width=.675\linewidth]{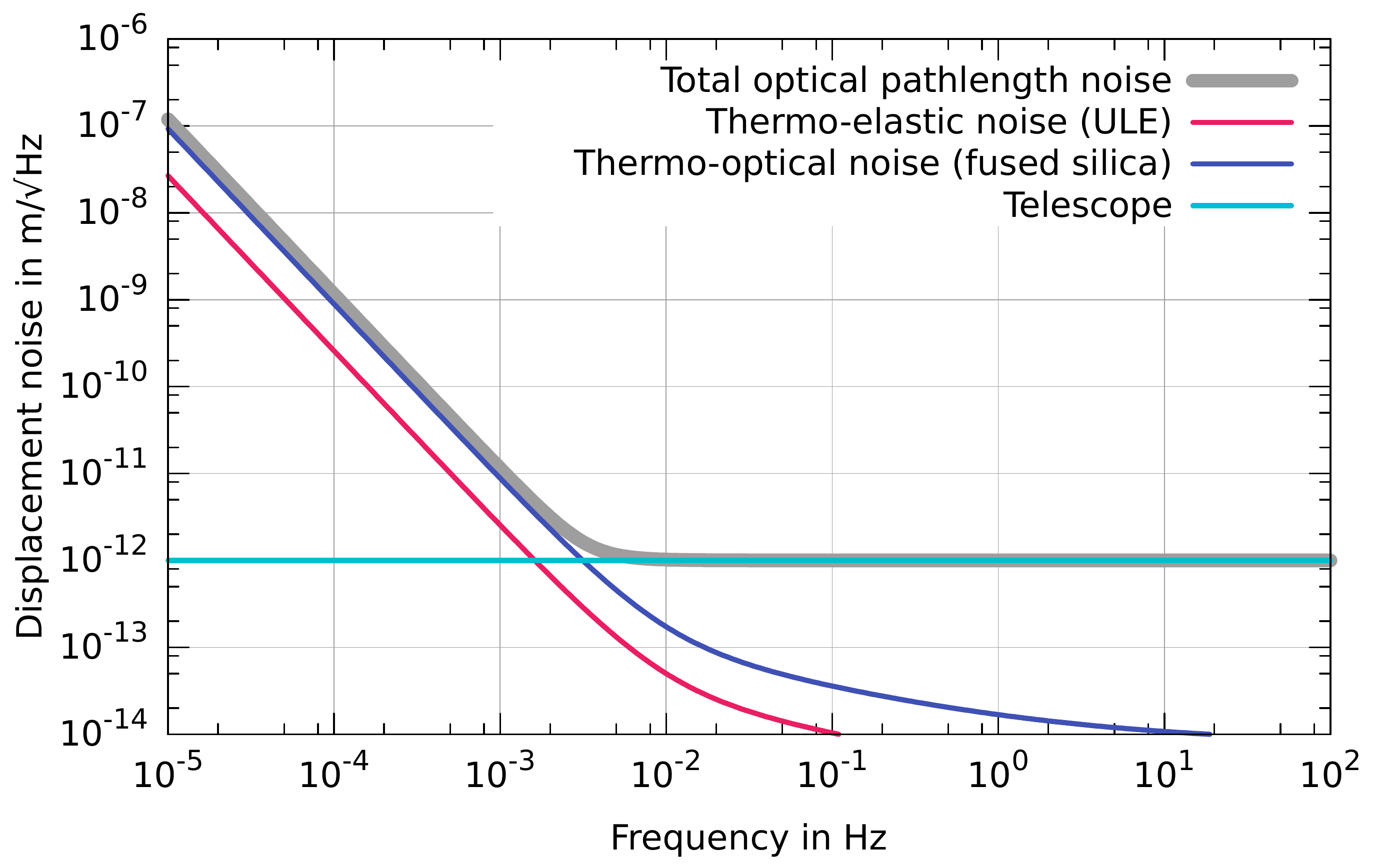}
	\caption{Total optical path length noise and individual contributions from optical bench and telescope. The summation of the contributions is described in Equation~(\ref{eq:xulefs}).}
	\label{fig:optical-pathlength-noise}
\end{figure}

While both path length noise contributions of the optical bench add linearly---since they are the result of the very same temperature fluctuations---the optical path length noise of the telescope relates to an uncorrelated temperature noise and hence adds quadratically. Thus the total optical path length noise has to be written as
\begin{equation}
\widetilde{x}_\text{opn}^\text{total}\left(f\right) = \sqrt{{\underbrace{\left[\widetilde{x}_\text{opn}^\text{ule}\left(f\right) + \widetilde{x}_\text{opn}^\text{fs}\left(f\right)\right]}_\text{\text{optical bench}}}^2 + {\underbrace{\left(\widetilde{x}_\text{opn}^{tel}\right)}_\text{\text{telescope}}}^2}~.
\label{eq:xulefs}
\end{equation}
All optical path length noise contributions and the total optical path length noise are plotted as a function of the Fourier frequency $f$ in Figure~\ref{fig:optical-pathlength-noise}.\\

Other sources of optical path length noise, like a non-uniform change in temperature, temperature gradients, and tilt-to-path length coupling \cite{schuster2014vanishing}, are neglected in this study. These contributions are either specific to the detailed mission design and hence hard to generalize, or based on complex coupling mechanisms and hence difficult to predict. 

\subsection{Acceleration Noise}

Residual forces on the proof masses, like Coulomb forces induced from imperfect cancellation of charges, surface effects, residual gas pressure, etc. result in an acceleration of the proof masses. We assume a white acceleration noise of $\num{3e-15}\,\text{m}\:\text{s}^{-2}/\sqrt{\text{Hz}}$ that can be described as frequency dependent displacement noise by
\begin{equation}
\widetilde{x}_\text{acc}\left(f\right) = \num{3e-15}\,\frac{\text{m}/\text{s}^2}{\sqrt{\text{Hz}}} \times \frac{1}{\left(2\pi\:f\right)^2} ~.
\label{eq:acceleration}
\end{equation}
In reality this function might be more complex due to the vast number of different effects acting on the proof masses. The above value is just a rough estimate and does not include, for example, a shift in the local gravitational field due to spacecraft position jitter. A more realistic acceleration noise model will be a direct heritage from the LISA Pathfinder mission \cite{anza2005ltp}, scheduled to launch in 2015.

\subsection{Metrology and Data Processing}

The individual inter-spacecraft interferometers  place one arm inside the spacecraft while the arm sensitive to gravitational waves is placed between spacecraft, which results in a huge arm length difference that is equal to the spacecraft separation distance $L_\text{arm}$. As in any unequal-arm Michelson interferometer, the laser frequency noise $\widetilde{\vartheta}_\text{pre}$ of the pre-stabilized laser at frequency $\vartheta = c/\lambda_\text{laser}$ directly translates to displacement noise with
\begin{equation}
\widetilde{x}_\text{ms}^\text{lfn} = L_\text{arm} \times \frac{\widetilde{\vartheta}_\text{pre}}{\vartheta} = \num{0.00206}\,\frac{\text{m}}{\sqrt{\text{Hz}}}~.
\label{eq:freqtodisp}
\end{equation}
This noise level dominates the entire observatory, but it can be suppressed  by a data post-processing technique called time-delay interferometry (TDI) \cite{otto2012tdi, tinto2014time}. Here, signals from different interferometers are time-shifted and combined in such a way that laser frequency noise cancels to the greatest extent. This only works if A) we read out all beat-notes in the heterodyne signal with sufficient precision, B) we have accurate knowledge of the inter-spacecraft separation distance, and C) we have precise time stamps of all measurements with respect to a constallation wide clock.
The latter information will be used to determine the correct time-shifts in post-processing. It is gained by a combination of
\begin{enumerate} 
	\item \textbf{spacecraft position triangulation} by the Deep Space Network,
	\item \textbf{ranging} with delayed pseudo random noise (PRN) codes modulated onto the laser beams \cite{esteban2009optical}, and
	\item \textbf{raw data pre-processing} by Kalman filters to recover the ranging information and base all measurements on a common reference frequency \cite{wang2014first}.
\end{enumerate}


Everything considered, we assume that the knowledge of the absolute spacecraft separation is better than $L_\text{ranging} = \num{0.1}$\,m. The amount of residual displacement noise due to laser frequency noise after TDI highly depends on this value hence we basically construct a virtual Michelson interferometer with an arm length difference equal to the ranging accuracy. We can thus calculate the equivalent displacement noise by simply adapting Equation~(\ref{eq:freqtodisp}) and get
\begin{equation}
\widetilde{x}_\text{ms}^\text{tdi} = L_\text{ranging} \times \frac{\widetilde{\vartheta}_\text{pre}}{\vartheta} = \num{1.03e-13}\,\frac{\text{m}}{\sqrt{\text{Hz}}}~.
\label{eq:ranging}
\end{equation}

On top of that we assume an ancillary phase error in the signal read-out of $\num{6}$\,$\upmu\text{rad}/{\sqrt{\text{Hz}}}$ at the maximum heterodyne frequency \cite{FinalReport}. This translates to a displacement noise equivalent of
\begin{equation}
\widetilde{x}_\text{ms}^\text{pm} = \num{1.016e-12}\,\frac{\text{m}}{\sqrt{\text{Hz}}}.
\label{eq:phasemeter}
\end{equation}

\begin{figure}[htb!]\centering
	\includegraphics[width=.675\linewidth]{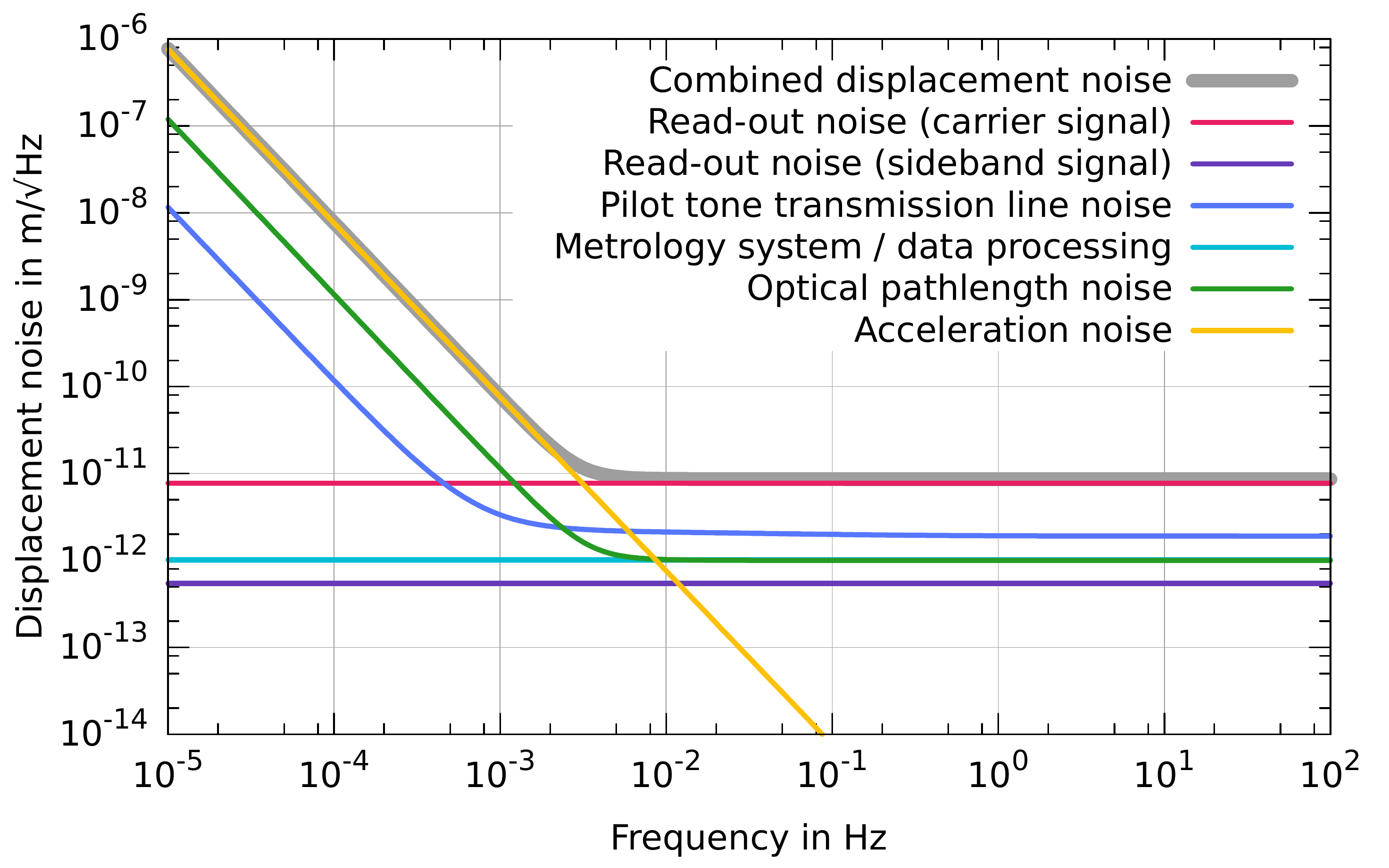}
	\caption{All effects that contribute to apparent displacement noise grouped into cetegories, and the resultant overall noise limit (combined displacement noise).}
	\label{fig:displacement-noise}
\end{figure}

While this read-out noise shows up in every single data stream, the ranging accuracy only comes into play when multiple links are combined. Technically speaking, each individual link is still limited by the noise level calculated in Equation~(\ref{eq:freqtodisp}). Nevertheless, for reasons of simplification, we add a metrology system and data processing noise level of
\begin{equation}
\widetilde{x}_\text{ms}^\text{total} = \sqrt{\left(\widetilde{x}_\text{ms}^\text{pm}\right)^2 + \left(\widetilde{x}_\text{ms}^\text{tdi}\right)^2} = \num{1.016e-12}\,\frac{\text{m}}{\sqrt{\text{Hz}}}
\label{eq:msnoise}
\end{equation}
to the total displacement noise of each link. In this way, we can compare all displacement noise contributions, summarized in Figure~\ref{fig:displacement-noise}, and determine the limiting influence.

The proof mass acceleration noise, $\widetilde{x}_\text{acc}$, is correlated between different links that share the same proof mass. All other displacement noise contributions, combined in
\begin{equation}
\begin{split}
\widetilde{x}_\text{idp} =~& \sqrt{{\left(\left<\widetilde{x}_\text{r/o}^\text{total}\right>_\text{carrier}\right)}^2 + {\left(\left<\widetilde{x}_\text{r/o}^\text{total}\right>_\text{sidebands}\right)}^2} \:\overline{\vphantom{\rule{0pt}{3.36ex}}\:}\:\overline{\vphantom{\rule{0pt}{3.36ex}}\:}\:\overline{\vphantom{\rule{0pt}{3.36ex}}\:}\\[1ex]
& \overline{\vphantom{\rule{0pt}{2.58ex}}\:}\:\overline{\vphantom{\rule{0pt}{2.58ex}}\:}\:\overline{\vphantom{\rule{0pt}{2.58ex}}\:}\:\overline{\vphantom{\rule{0pt}{2.58ex}}+ {\left(\widetilde{x}_\text{tml}^\text{total}\right)}^2 + {\left(\widetilde{x}_\text{ms}^\text{total}\right)}^2 + {\left(\widetilde{x}_\text{opn}^\text{total}\right)}^2}~,
\end{split}
\label{eq:indnoise}
\end{equation}
are independent between links. The total displacement noise which is used in all further evaluation of the observatory's sensitivity is given by
\begin{equation}
\widetilde{x}_\text{total} = \sqrt{{\left(\widetilde{x}_\text{acc}\right)}^2 + {\left(\widetilde{x}_\text{idp}\right)}^2}~.
\label{eq:totalnoise}
\end{equation}

Table~\ref{tab:basic} lists all parameters that were used to deduce the total displacement noise.
While many details are still under investigation, these parameters correspond to values currently assumed to be likely applied to the actual 2034 ESA mission.
All parameters can be individually changed in the web application. 

\begin{table}
	\caption{\label{tab:basic}Parameters for the laser interferometric gravitational wave observatory investigated in this study that were used to deduce the total equivalent displacement noise and observatory sensitivity. These parameters correspond to values currently assumed to be likely applied to a space mission to be launched in 2034 by the European Space Agency. All parameters can be individually changed in the associated web application. }
	\begin{indented}
		\lineup	
		\item[]\begin{tabulary}{\linewidth}{LRL}
			
			\br
			\textbf{Parameter} & ~ & ~\textbf{Value}\\
			\mr
			Number of links  & $N_\text{links} =  $ &  ~ $ \num{6} $  ~\\
			Average arm length  & $L_\text{arm} = $ &  ~  $ \num{2000000} $  km  \\
			Heterodyne frequency (max.)  & $f_\text{het} =  $ &  ~  $ \num{18}$  MHz  \\
			Laser wavelength & $\lambda_\text{laser} =  $ & ~ $ \num{1064}$  nm \\
			Optical power (to telescope) & $ P_\text{tel} =  $ & ~ $ \num{1.65}$  W \\
			Relative intensity noise (laser) & $ \text{RIN} =  $ & ~ $ \num{1e-8}$  $/{\sqrt{\text{Hz}}}$ \\
			Laser frequency noise after pre-stabilization & $ \widetilde{\vartheta}_\text{pre}  =  $ & ~ $\num{290}$  $\text{Hz}/{\sqrt{\text{Hz}}}$ \\
			Telescope diameter & $ d_\text{tel} =  $ & ~ $ \num{26} $  cm \\
			Optical efficiency (receive path) & $ \eta_\text{opt} =  $ & ~ $ \num{70} $  \% \\
			Beam waist position$^{\rm 1}$ &  \multicolumn{2}{c}{at transmitting telescope}  \\
			Optimum beam waist$^{\rm 1}$ & $ \omega_0 =  $ & ~ $ \num{11.6} $ cm \\
			Received laser power$^{\rm 1}$ & $ P_\text{rec} =  $ & ~ $ \num{836.61} $ pW \\
			Local laser power$^{\rm 1}$ & $P_\text{local} =  $ & ~ $ \num{1.75e-3} $ W ~\\
			Temperature noise at electronics and electro-optics & $\widetilde{T}_\text{el}\left(f\right) \hphantom{=}  $ & 
			~ see Section~\ref{sec:temperature}~\\
			Temperature noise at optical bench & $\widetilde{T}_\text{ob}\left(f\right) \hphantom{=}  $ & ~ see Section~\ref{sec:temperature} ~\\
			Photodiodes & $ N_\text{pd} =  $ & ~ $ 4 $ segments \\
			Quantum efficiency of photodiodes& $ \eta_\text{pd} =  $ & ~ $ \num{80} $ \% \\
			Photodiode responsivity& $ R_\text{pd} =  $ & ~ $ \num{0.69} $ $\text{A}/\text{W}$ \\
			Current noise (photodiode)& $ \widetilde{I}_\text{pd} =  $ & ~ $ \num{2} $ $\text{pA}/{\sqrt{\text{Hz}}}$ \\
			Capacitance (photodiode)& $ C_\text{pd} =  $ & ~ $ \num{10} $ pF \\
			Voltage noise (transimpedance amplifier)& $ \widetilde{U}_\text{pd} =  $ & ~ $ \num{2} $ $\text{nV}/{\sqrt{\text{Hz}}}$ \\
			Heterodyne efficiency & $ \eta_\text{het} =  $ & ~ $ \num{70} $  \% \\
			Single first-order sideband power (in parts of carrier power) & $ \frac{\text{sideband}}{\text{carrier}} =  $ & ~ $ \num{7.5} $ \% ~\\
			Modulation frequency & $f_\text{mod} =  $ & ~ $ \num{2.4} $\,GHz \\
			Timing jitter (electronics) & $\widetilde{t}_\text{el} =  $ & ~ $ \num{4e-14} $\,$\text{s}/{\sqrt{\text{Hz}}}$  \\
			Thermal stability (cables) & $\left(\frac{\delta \phi}{\delta T}\right)_\text{cables} = $ & ~ $ \num{7} $\,$\text{mrad}/{\left(\text{K}\:\text{m}\:\text{GHz}\right)}$  \\
			Thermal stability (fibers) & $\left(\frac{\delta \phi}{\delta T}\right)_\text{fibers} =  $ & ~ $ \num{1}$\,$\text{mrad}/{\left(\text{K}\:\text{m}\:\text{GHz}\right)}$ \\
			Total length (cables) & $l_\text{cables} =  $ & ~ $ \num{2} $\,m  \\
			Total length (fibers) & $l_\text{fibers} =  $ & ~ $ \num{5} $\,m  \\
			Noise (EOM) & $\widetilde{x}_\text{tml}^\text{eom} =  $ & ~ $ \num{3.81e-13} $\,$\text{m}/{\sqrt{\text{Hz}}}$  \\
			Noise (fiber amplifier) & $\widetilde{x}_\text{tml}^\text{fa} = $ & ~ $ \num{7.62e-13} $\,$\text{m}/{\sqrt{\text{Hz}}}$  \\
			Optical path length difference (in fused silica) & $ \text{OPD}_\text{fs} =  $ & ~ $ \num{29} $  mm \\
			Optical path length difference (on optical bench) & $ \text{OPD}_\text{ob} =  $ & ~ $ \num{565} $  mm \\
			Optical path length noise (telescope) & $ \widetilde{x}_\text{opn}^{tel} =  $ & ~ $ \num{1} $ $\text{pm}/{\sqrt{\text{Hz}}}$ \\
			Ranging accuracy (rms)& $ L_\text{ranging} =  $ & ~ $ \num{0.1} $  m \\
			Acceleration noise & $ \widetilde{x}_\text{acc}\left(f\right) =  $ & ~ $ \num{3e-15} $  $\frac{\text{m}/\text{s}^2}{\sqrt{\text{Hz}}} \times \frac{1}{\left(2\pi\:f\right)^2}$ \\
			Metrology system read-out noise & $ \widetilde{x}_\text{ms}^\text{pm} =  $ & ~ $ \num{1.016e-12} $  $\text{m}/{\sqrt{\text{Hz}}}$ \\
			\br
		\end{tabulary}
		\item[] $^{\rm 1}$ Values were optimized automatically.
	\end{indented}
\end{table}

\section{Observatory Sensitivity}

\begin{figure}[htb!]\centering
	\includegraphics[width=.675\linewidth]{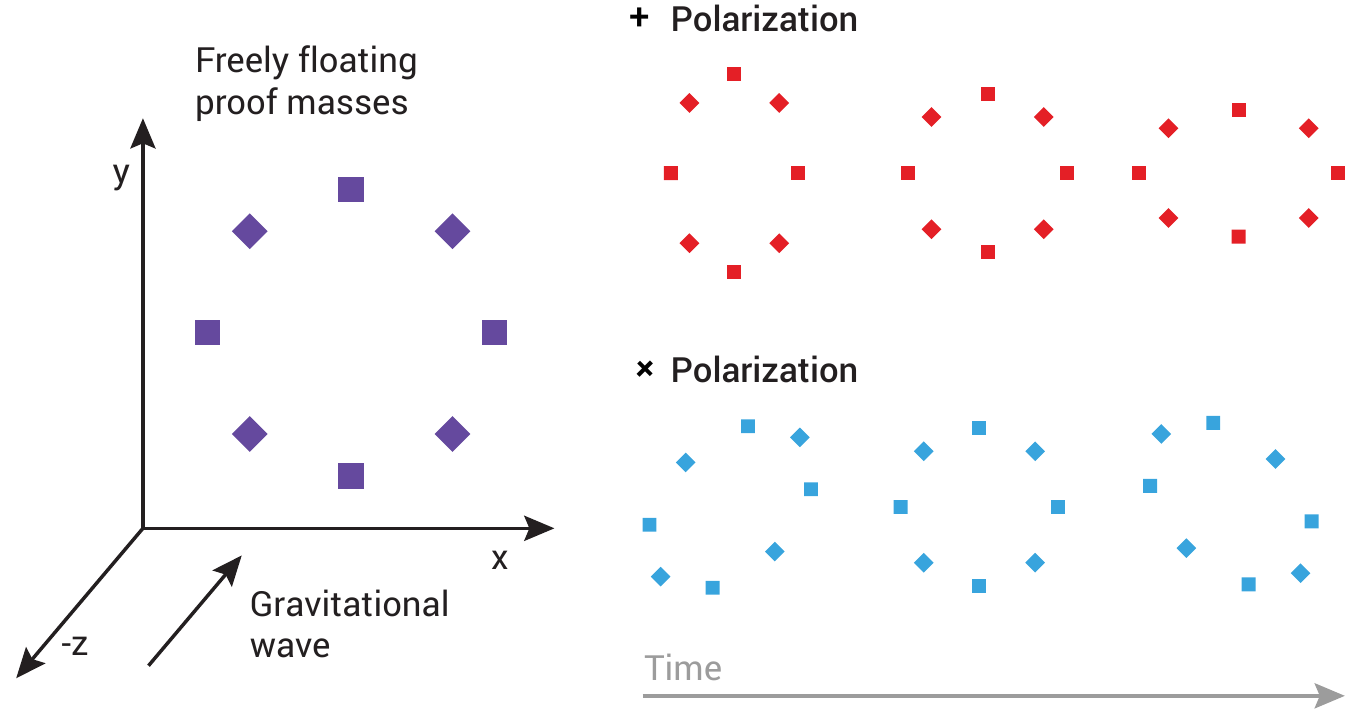}
	\caption{A ring of proof masses freely floating in the xy-plane and a gravitational wave that propagates along the z-direction. While a $+$-polarized wave will change the proper distance in x and y directions, the influence of a $\times$-polarized wave is rotated by 45$^{\circ}$ so that distances along the x- and y-axis remain unaffected.}
	\label{fig:polarization}
\end{figure}

Gravitational waves stretch and compress spacetime perpendicular to the direction of travel and cause directly observable distance fluctuations between proof masses.  Let's assume we have a ring of cubes freely floating in the xy-plane and a gravitational wave propagates along the z-direction. As illustrated in Figure~\ref{fig:polarization}, the distance between the masses oscillates with time, the direction of this oscillation depends on the polarization of the gravitational wave. The usual basic set of polarization states are plus ($+$) and cross ($\times$) polarization, others can be formed by linear combinations of these two.

\subsection{Single Link}
\begin{figure}[htb!]\centering
	\includegraphics[width=.675\linewidth]{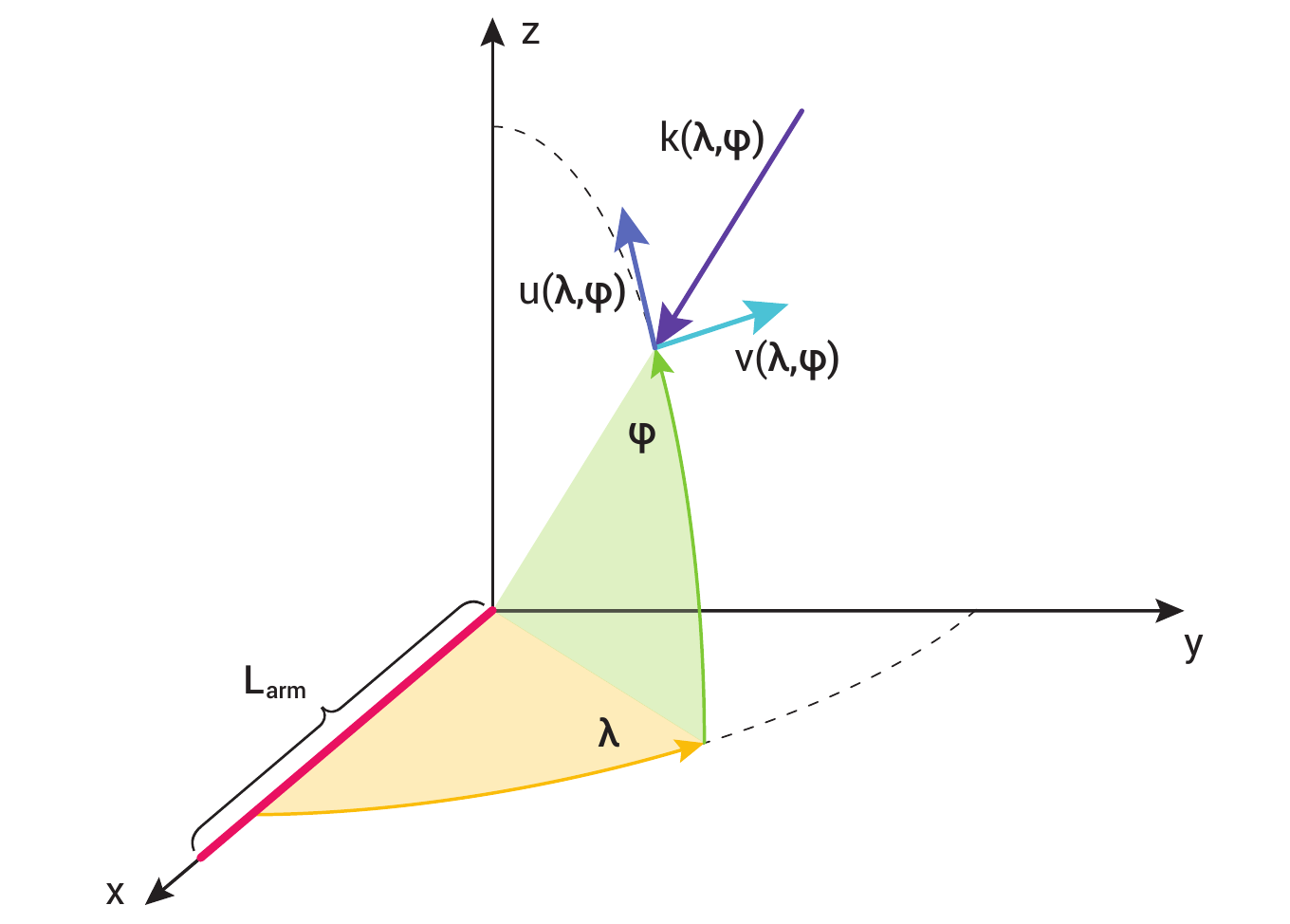}
	\caption{The response to gravitational waves of a single link (here: aligned with the x-axis) depends on the gravitational wave incident vector $k$ with orthogonal components $u$ and $v$. The actual oscillation is polarization dependent as indicated in  Figure~\ref{fig:polarization}.}
	\label{fig:polarkuv}
\end{figure}

To calculate the impact on one link when a gravitational wave passes though the observatory, we align the link with the the unit vector $\bm{e}_\text{x}$ (in the direction of the x-axis) and observe a gravitational wave that propagates along a vector $\bm{k}\left(\phi,\lambda\right) = - \left(\cos{\phi} \cos{\lambda}, \cos{\phi} \sin{\lambda}, \sin{\phi} \right)$. The use of polar coordinates with latitude $\phi$ and longitude $\lambda$ is illustrated in Figure~\ref{fig:polarkuv}.
The oscillation of spacetime transverse to $\bm{k}$ happens along the orthogonal unit vectors $\bm{u}\left(\phi,\lambda\right)$ and $\bm{v}\left(\phi,\lambda\right)$ with x-axis components $\bm{u} \times \bm{e}_\text{x}= \sin{\phi} \times \cos{\lambda}$ and $\bm{v} \times \bm{e}_\text{x} = \sin{\lambda}$. Two influences have to be considered both of which can reduce the impact of a gravitational wave of the link: the antenna pattern and the frequency response.\\

The antenna pattern $F\left(\lambda, \phi\right)$ is a function of the sky position of the source (vector $\bm{k}$) and combines the response for both polarization states.
For a single link aligned with the x-axis it can be expressed by
\begin{equation}
\begin{split}
F\left(\lambda, \phi\right) ~&= \frac{1}{2}\Big[\underbrace{\left(\bm{u}\times\bm{e}_\text{x}\right)^2-\left(\bm{v}\times\bm{e}_\text{x}\right)^2}_{+~\text{polarization}} + \underbrace{2\left(\bm{u}\times\bm{e}_\text{x}\right)\left(\bm{v}\times\bm{e}_\text{x}\right)}_{\times~\text{polarization}}\Big]\\
&= \frac{1}{2}\:\left(\sin^2\phi \cos^2\lambda - \sin^2\lambda + 2 \sin\phi \cos\lambda \sin\lambda\right) ~.
\end{split}
\label{eq:antenna}
\end{equation}
This function basically indicates which directions the gravitational wave observatory is sensitive to. While the link will not be influenced by gravitational waves propagating along the x-axis at all, independent of the polarization, the maximum impact can be observed for a $+$-polarized gravitational wave propagation orthogonal to the x-axis. A $\times$-polarized wave however does have no effect on the x-axis if propagating orthogonal to the x-axis. In general, laser interferometric gravitational wave observatories are sensitive to a very large fraction of the sky, hence they are usually referred to as omni-directional detectors.\\

The frequency response $R\left(f, \lambda, \phi\right)$ is a function of the gravitational wave frequency $f$, or---more accurately---the frequency of the influence of the gravitational wave propagating along vector $\bm{k}$ projected on the link vector $\bm{x}$. It can be expressed by
\begin{equation}
R\left(f, \lambda, \phi\right) = \frac{e^{2\pi\:i\:\left[1-\bm{k}\bm{x}\right]\:L_\text{arm}/\frac{c}{f}}-1}{2\pi\:i\:\left[1-\bm{k}\bm{x}\right]\:L_\text{arm}/\frac{c}{f}} \times e^{-2\pi\:i\:\bm{k}}
\label{eq:responseB}
\end{equation}
and depends on the actual arm length in relation to the wavelength of the gravitational wave $L_\text{arm}/\frac{c}{f}$.
At low frequencies the frequency response is flat. For high frequencies, when the projected wavelength equals a multiple of the arm length, the effect of the gravitational wave oscillation cancels out and the sensitivity is reduced.\\

Both influences combined give the total single link transfer function
\begin{equation}
T_\text{link}\left(f, \lambda, \phi\right) = F\left(\lambda, \phi\right) \times R\left(f, \lambda, \phi\right)
\label{eq:transferfunction}
\end{equation}
and we can calculate its absolute average value over all sky positions ($\lambda=0 \dots 2\pi$, $\phi=-\pi/2 \dots \pi/2$)
\begin{equation}
T_\text{link}\left(f\right) = \sqrt{\left< \left|T\left(f, \lambda, \phi\right)\right|^2\right>_\text{\text{sky}}}~.
\label{eq:transferfunctionsky}
\end{equation}
The effective strain sensitivity for a single link can now be formulated as the displacement noise over the single link transfer function
\begin{equation}
\sqrt{S_n\left(f\right)_\text{link}} = \frac{\widetilde{x}_\text{total}}{T_\text{link}\left(f\right) \times L_\text{arm}}~.
\label{eq:transferfunctionskyB}
\end{equation}
It is given in relative units ($\text{m}/\sqrt{\text{Hz}}$ per meter $= 1/\sqrt{\text{Hz}}$), thus the division by the arm length $L_\text{arm}$.\\

\begin{figure}[htb!]\centering
	\includegraphics[width=.675\linewidth]{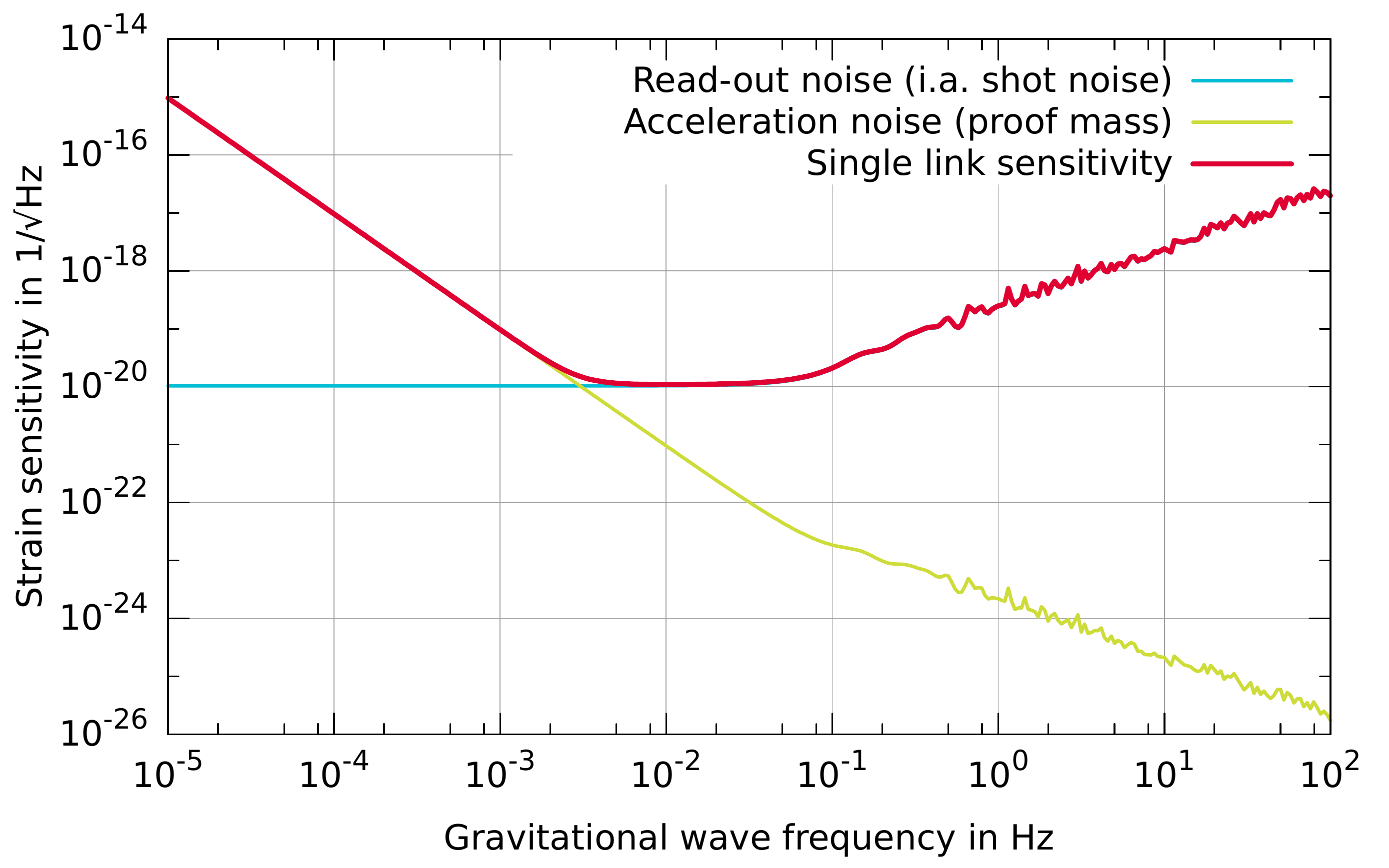}
	\caption{Single link strain sensitivity for the gravitational wave observatory specified in this study compared to the individual noise contributions by carrier signal read-out noise and proof mass acceleration noise.}
	\label{fig:single-arm-noise}
\end{figure}

Figure~\ref{fig:single-arm-noise} shows the effective single link strain sensitivity the observatory specified above (red trace). Individual contributions by carrier signal read-out noise (blue) and proof mass acceleration noise (green) are shown. In a carefully designed observatory these two influences should limit the overall sensitivity.

The wiggles observable in the reduced sensitivity at high frequencies result from an attempt to reduce the response time of the web application---ideally below \num{400}\,ms, known as the Doherty threshold \cite{doherty1982economic}---and the load on the web server performing the calculations. Thus we chose a sloppy averaging over only four values for $\lambda = \left[ 0, 2, 4, 6\right]$ and four values for $\phi = \left[ -1.41, 0.47, 0.47, 1.41 \right]$. Yet this alone accounts for 16 different transfer functions with $> 300$ values each (50 values per frequency decade). For a perfect average over all sky positions the slope at high frequencies should become continuous.

\subsection{Full Observatory}

The single link sensitivity is a good indicator of the observatory's performance. It can be used to compare different sets of parameters that share the same constellation to quickly identify limiting noise sources. This is the main purpose of the developed web application. In reality though, contributions like sideband signal read-out noise or pilot tone transmission chain noise have no effect when considering only one link. Instead, the sensitivity would be substantially reduced by frequency noise of the pre-stabilized lasers. Hence a single link cannot be used to detect gravitational waves.

To calculate the actual sensitivity of the full observatory, we have to consider the combined responsivity of all links including their individual spatial orientation. Within the ranging accuracy,  signals must be precisely time-shifted to compensate for laser frequency noise in a TDI simulation with realistic input data streams. On top of that, all data has to be referenced to a common frequency considering the pilot tone transmission fidelity. This process is described  in-depth by \cite{tinto2002time, otto2012tdi, lrr-2014-6}, but would require too much resources within the scope of the web application.\\

\begin{figure}[htb!]\centering
	\includegraphics[width=.675\linewidth]{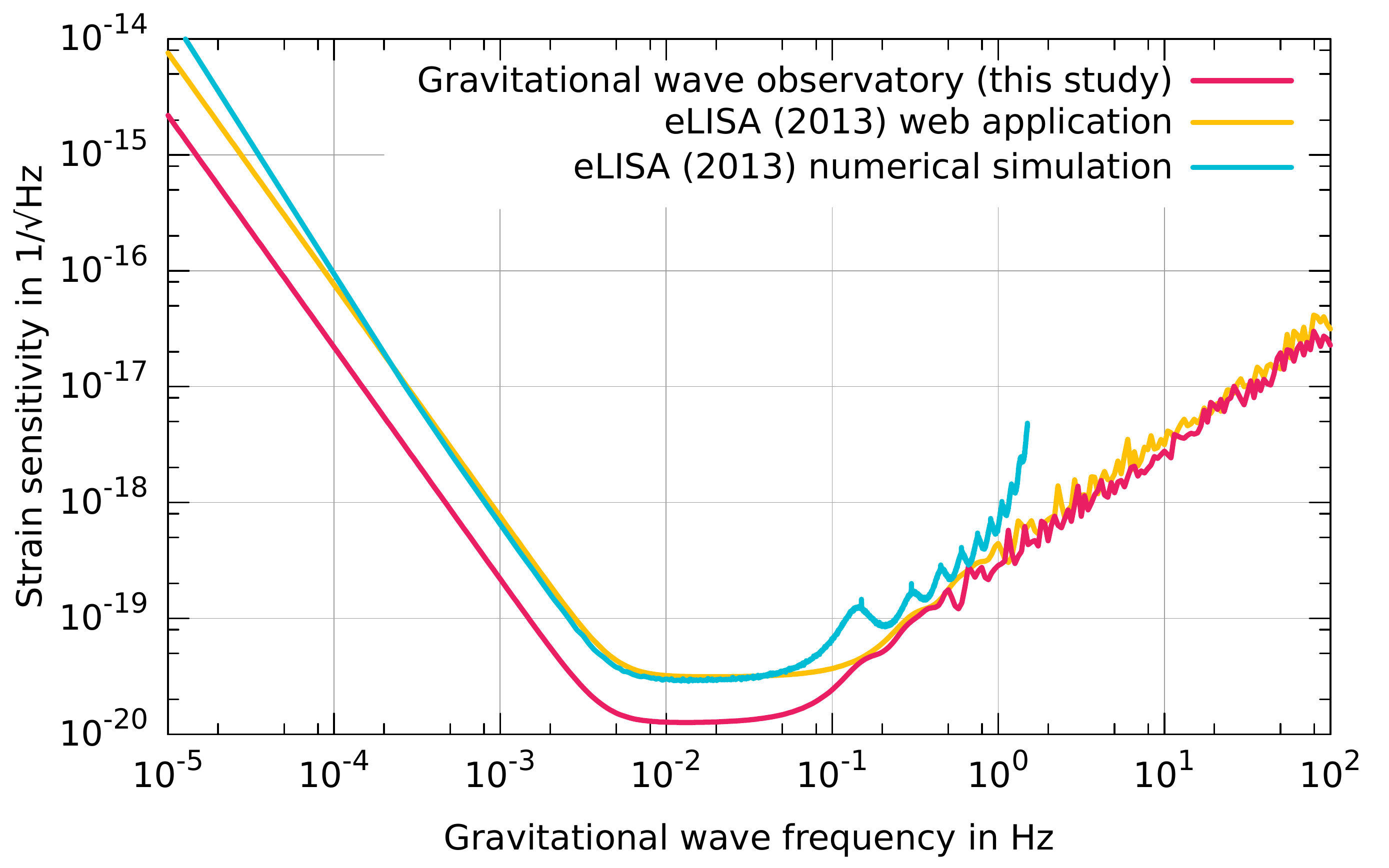}
	\caption{Appoximate total strain sensitivity (all sky and polarization average) for the described eLISA-like observatory compared to a numerical TDI simulation for the eLISA (2013) gravitational wave observatory.}
	\label{fig:overview-noise}
\end{figure}

A good estimate of the full observatory sensitivity without excessive computational effort can be extrapolated from the single link sensitivity since in our case it already contains noise contributions due to limited ranging accuracy and pilot tone transmission fidelity. There are two effects: A) The combination of time-shifted signals results in an increased noise level: a thorough study of \cite{lrr-2014-6, cornish2001detecting} reveals that for a 60$^{\circ}$ virtual Michelson interferometer, TDI increases the proof mass acceleration noise at low frequencies by a factor of 4, while all other displacement noise contributions---which are uncorrelated between links---are increased by a factor of 2. B) The total number of virtual Michelson interferometers results in a general sensitivity improvement: a 3-arm triangular observatory can form three individual virtual Michelson interferometers, hence the overall sensitivity increases roughly by a factor of $\sqrt{3}$. Accordingly we can write the full observatory strain sensitivity approximately\footnote{This approximation is not valid for octahedral (24 link) configurations where an enhanced post-processing technique called displacement-noise free interferometry (DFI) \cite{chen2006interferometers, wang2013octahedron} is used to suppress proof mass acceleration noise alongside any other spacecraft common mode displacement noise as well as laser frequency noise.}  as 
\begin{equation}
\sqrt{S_n\left(f\right)_\text{obs}} \approx \frac{1}{\sqrt{3}} \times \frac{\sqrt{\left(4 \times \widetilde{x}_\text{acc}\right)^2 + \left(2 \times \widetilde{x}_\text{idp}\right)^2}}{T_\text{link}\left(f\right) \times L_\text{arm}}~.
\label{eq:transferfunctionobs}
\end{equation}
Figure~\ref{fig:overview-noise} shows this full observatory sensitivity in red.

For comparison a numerical TDI simulation that was done for the eLISA (2013) gravitational wave observatory mission study as part of `The Gravitational Universe' White Paper \cite{seoane2013gravitational} is shown in blue. eLISA (2013) used slightly different parameters, namely only 4 links, smaller arm length, telescope diameter and heterodyne frequency, and higher laser power. A list of all parameters that differ from the ones in this study can be found in Table~\ref{tab:elisa}.

\begin{table}[htb!]
	\caption{\label{tab:elisa}Parameters that differ from Table~\ref{tab:basic} to correspond to the parameter set used for the eLISA (2013) mission study.}
	\begin{indented}
		\lineup	
		\item[]\begin{tabulary}{\linewidth}{LRL}
			
			\br
			\textbf{Parameter} & ~ & ~\textbf{Value}\\
			\mr
			Number of links  ~ ~ & $N_\text{links} =  $ &  ~ $ \num{4} $  ~\\
			Average arm length ~ ~  & $L_\text{arm} = $ &  ~  $ \num{1000000} $  km  \\
			Heterodyne frequency (max.)  ~ ~ & $f_\text{het} =  $ &  ~  $ \num{12}$  MHz  \\
			Optical power (to telescope) ~ ~ & $ P_\text{tel} =  $ & ~ $ \num{2}$  W \\
			Telescope diameter ~ ~ & $ d_\text{tel} =  $ & ~ $ \num{20} $  cm \\
			\br
		\end{tabulary}
	\end{indented}
\end{table}

The result from the web application for this new parameter set with the full observatory strain sensitivity approximated by Equation~(\ref{eq:transferfunctionobs}) is shown in orange. Although for this approximated sensitivity the wiggles at high frequencies are again due to a sloppy averaging, similar wiggles in the sensitivity deduced by the numerical simulation are a real consequence of the TDI algorithms. This shows the limitations of our approximation. Nevertheless it is sufficient for the purpose of parameter optimization is a very close match to the real sensitivity. Thus we can use it to investigate the astrophysical relevance of the observatory.

\subsection{Astrophysical Sources}

\begin{figure}[htb!]\centering
	\includegraphics[width=\linewidth]{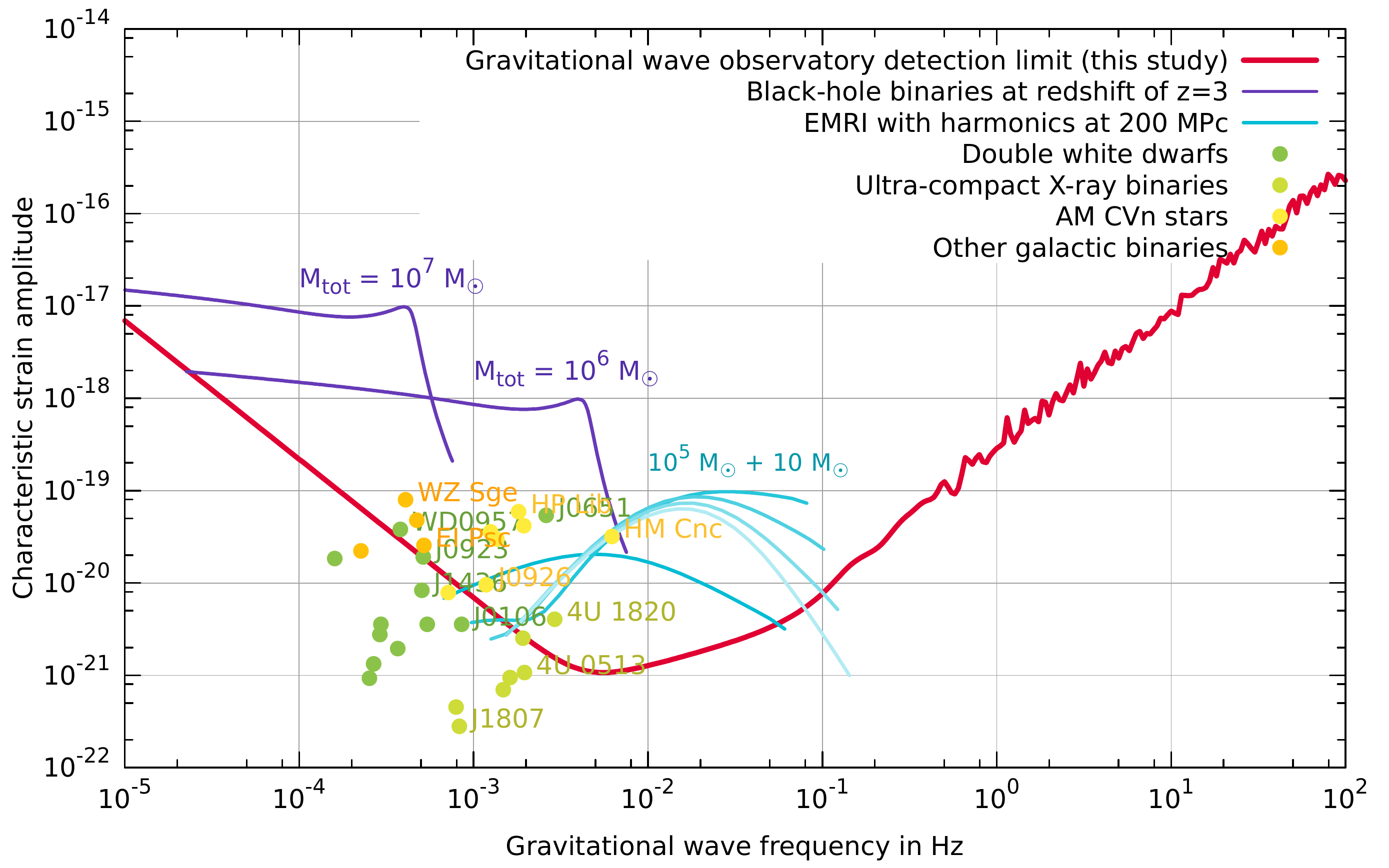}
	\caption{Observatory detection limit (for SNR$=1$) and dimensionless characteristic strain amplitudes for different gravitational wave sources. Two traces for systems of binary black holes at redshift of $z=3$ (total mass $M_\text{tot}=10^7 M_{\odot}$ and $=10^6 M_{\odot}$), where the former trace starts at low frequencies $\approx 1$ month, the latter $\approx 1$ year before the plunge (spike in the trace).
		First 5 harmonics of one eccentric Extreme Mass Ratio Inspiral (EMRI) for an object with mass $m=10 M_{\odot}$ captured by a massive black hole of mass $M=10^5 M_{\odot}$ at 200\,Mpc distance. The EMRI trace starts at low frequencies many years before the merger. A selection of known ultra-compact binary stars (dots) for 1 year of observation time.
	}
	\label{fig:amplitude-noise}
\end{figure}

The scientific value of an observatory is related to the number and type of sources it can detect. In Figure~\ref{fig:amplitude-noise} we use all parameters of this study (see Table~\ref{tab:basic}) to plot the
observatory's detection limit
\begin{equation}
h_\text{c}\left(f\right) = \sqrt{f} \times \sqrt{S_n\left(f\right)_\text{obs}}
\label{eq:transferfunctionobs2}
\end{equation}
where the signal-to-noise ratio equals 1. We can compare this to the characteristic gravitational wave strain amplitudes (given in $\text{m}/\text{m}$) for selected gravitational wave sources. For quasi-monochromatic sources the accumulated signal after one year of observation time is given. Amplitudes of all other broadband sources are plotted as is, although their actual SNR can be higher due to matched filtering techniques during data analysis.

There are three categories of astrophysical phenomena that are known to emit gravitational waves at frequencies and amplitudes accessible to laser interferometric observatories in space.

\begin{enumerate} 
	\item \textbf{Massive black hole binaries}: the coalescence of two supermassive black holes.
	\item \textbf{Extreme Mass Ratio Inspirals (EMRIs)}: a compact star or stellar mass black hole captured in a highly relativistic orbit around a massive black hole.
	\item \textbf{Ultra-compact binaries}: systems of white dwarfs, neutron stars, or stellar mass black holes in tight orbit.
\end{enumerate}

The amount of energy emitted in form of gravitational waves is very different between these phenomena. Thus the distance to detectable sources varies greatly.\\

There may also be gravitational wave signals of yet unknown origin within the sensitivity of the described observatory. It must be remembered that no one ever detected signals in this frequency range and new discoveries that radically expand our knowledge of fundamental physics and astrophysical processes are most likely.

\subsubsection{Massive Black Hole Binaries}

Galaxies usually harbor one or more massive central black holes, some million times heavier than our Sun. When galaxies coalesce, these black holes will merge eventually, releasing huge amounts of gravitational radiation in the process. Signals should be easily detectable for redshifts of $z=3$ and higher (at a distance of over $\approx 22$ billion light years) even many months before the final plunge. Such gravitational waves originated over 12 billion years ago, so we can basically detect such events throughout the entire observable universe.

Figure~\ref{fig:amplitude-noise} shows two examples taken from \cite{antoine}. In each case systems of two massive black holes at redshift of $z=3$ are shown, one with a total mass $M_\text{tot}=10^7 M_{\odot}$, the other with $M_\text{tot}=10^6 M_{\odot}$. While the former signal starts at low frequencies approximately one month before the plunge (spike in the trace), the latter signal is shown for the final year before plunge. The detection of such signals will reveal the masses and spins of the two black holes, and shed light on the evolution and merger history of galaxies all the way back to shortly after the Big Bang.

\subsubsection{Extreme Mass Ratio Inspirals (EMRIs)}

Compact stars or stellar mass black holes can be captured by the massive central black holes of galaxies. They are spiraling through the strongest gravitational field regions just a few Schwarzschild radii from the event horizon \cite{seoane2013gravitational}. Such events should be resolvable many years before the merger for sources at hundreds of MPc distance. This corresponds to $\approx 2$ billion light years and easily contains the entire Laniakea Supercluster and all neighboring structures, accumulating signals from over 500 million galaxies \cite{powell2006atlas}.

The highly relativistic orbits result in feature-rich waveforms with many harmonics. Figure~\ref{fig:amplitude-noise} shows the first 5 harmonics of an eccentric EMRI for an object with mass $m=10 M_{\odot}$ captured by a massive black hole of mass $M=10^5 M_{\odot}$ at 200\,Mpc distance \cite{alberto}. The detection of such signals will allow a deep view into galactic nuclei for the very first time.

\subsubsection{Ultra-compact Binaries}
About half of the stars in the Milky Way are thought to exist in binary systems \cite{horch2014most}, some times even in orbits so compact that orbital periods are shorter than one hour. 
A list of all currently known ultra-compact binaries can be found in \cite{LISAwiki}. For many of these systems, parameters (orbital period, distance, and masses) are known with sufficient accuracy so we can calculate an order-of-magnitude gravitational wave signal prediction. Following \cite{LectureA} we find the dimensionless gravitational wave strain amplitude measured at a distance $d$ from the source within one orbital frequency bin to be
\begin{equation}
h_\text{c} = 2\left(4\pi\right)^{1/3} \times \frac{G^{5/3}}{c^4} f^{2/3}  m\times M^{2/3} \times \frac{1}{d}~, 
\label{eq:binaryamplitude}
\end{equation}
with $M=m_1+m_2$ being the total mass and $m=\frac{m_1 \times m_2}{m_1+m_2}$ the effective inertial mass. The frequency of the gravitational waves $f=2\times1/T$ is twice the orbital frequency and $G$ is the gravitational constant. 

All known ultra-compact binaries are quasi-monochromatic so they do not chirp appreciably during an observation of realistic length $T_\text{obs}$. Thus the frequency can be assumed to be constant over the mission duration $< T_\text{obs}$ and the signal amplitude accumulates to
\begin{equation}
h_\text{c}^\text{obs} = h_\text{c} \times \sqrt{N_\text{cycles}}~.
\label{eq:binaryamplitudeyear}
\end{equation}
Here $N_\text{cycles}=f\times T_\text{obs}$ depicts the number of cycles observable within the observation time. Figure~\ref{fig:amplitude-noise} shows all ultra-compact binaries for $T_\text{obs} = 1$~year.

We can observe Double white dwarf (WD) stars, ultra-compact X-ray binaries, AM Canum Venaticorum (AM CVn) stars, as well as any other cataclysmic variable (CV) stars, subdwarf B + WD binaries or double neutron stars out to distances of thousands of Pc. This corresponds to $\approx 30$ thousand light years and encloses our quadrant of the Milky Way galaxy with $\approx 50$ billion stars.\\

On top of that, there will be a noise contribution from the vast number of weak galactic binaries where individual sources cannot be disentangled in the data stream. The calculation of this noise usually involves a simulated catalog of millions of sources to find out how many sources are identifiable and which ones contribute to the overall noise floor, depending on the particular detector sensitivity. This simulation is not yet integrated in the developed web application and hence the confusion noise is not shown in Figure~\ref{fig:amplitude-noise}.

\section*{Web Application}
\addcontentsline{toc}{section}{Web application} 

\begin{figure}[htb!]\centering
	\includegraphics[width=\linewidth]{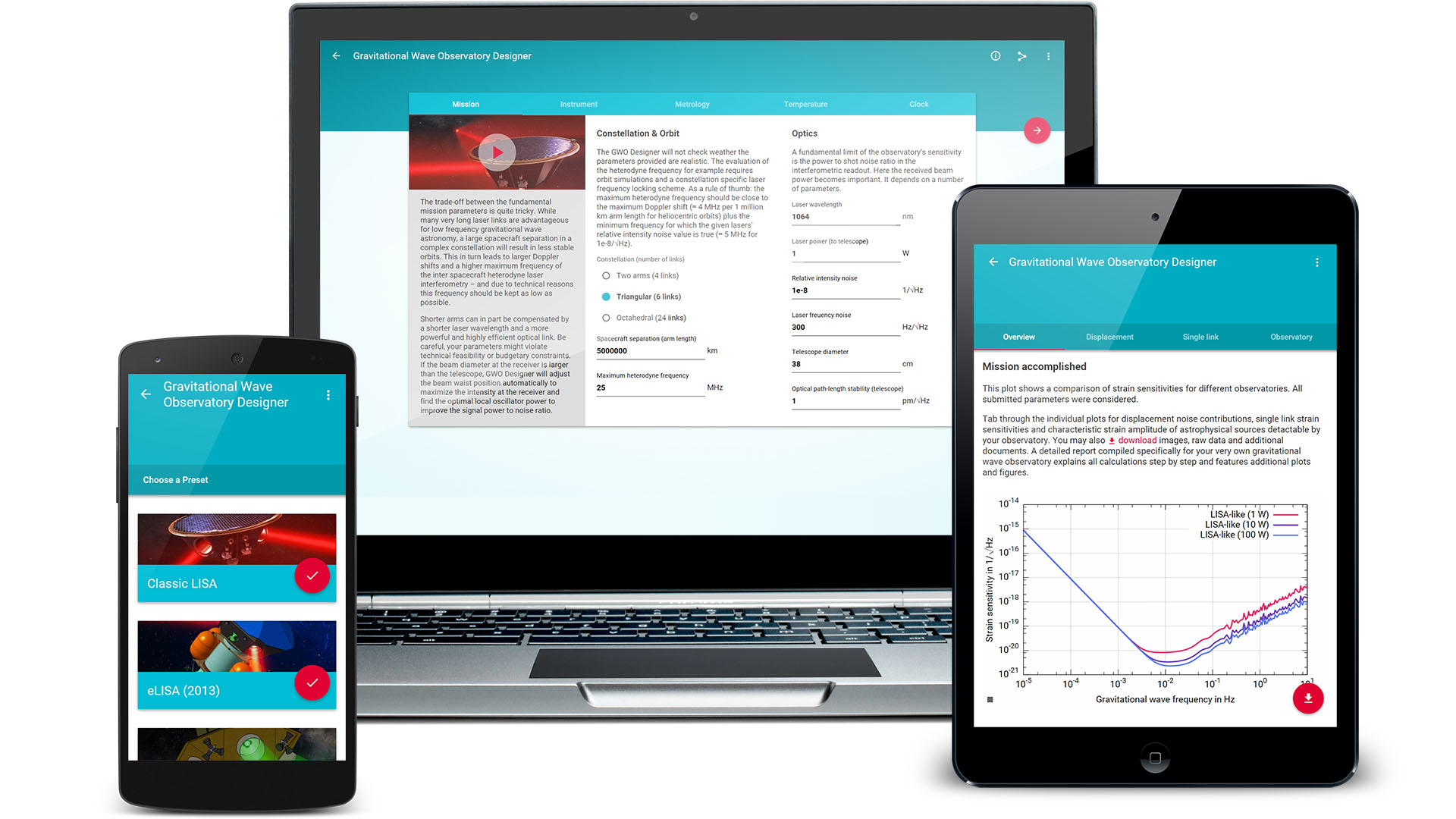}
				\caption{Graphical user interface of the ``Gravitational Wave Observatory Designer'': The compliance with Google Inc.'s `Material Design' guidelines allows for a unified user experience across a wide range of devices, screen sizes, and formats.}
				\label{fig:material}
	\end{figure}

All of the above calculations can be performed and documented for your specific set of mission parameters by the ``Gravitational Wave Observatory Designer''. This web application---which is publicly available on the Internet---was developed in the context of this study. It provides an HTML5 based graphical user interface (GUI) designed with jQuery, a cross-platform JavaScript library, and Elements from Polymer, an open-source Web Components-based library made available by Google Inc. Although only Chrome (and other Blink-based browsers like Opera) ship with native platform support for Web Components, a JavaScript foundation layer provides compatibility for the latest version of all `evergreen' (self updating) web browsers. That currently includes Chrome (also Android and Canary versions), Firefox, Internet Explorer (version 10 and up), and Safari (version 6 and up, also mobile versions). The compliance with Google Inc.'s `Material Design' guidelines allows for a unified user experience across a wide range of devices, screen sizes, and formats. Examples are shown in Figure~\ref{fig:material}.\\

All calculations are done by a Perl CGI back end that is connected to the GUI via Ajax, a technique for asynchronous client-side JavaScript and XML. It utilizes Perl modules such as Math::Cephes, PDL, and Math::Complex, and interfaces with gnuplot, an open source command-line program to generate graphics in various formats including interactive SVG plots. PDF documents are created by LaTeX, a document preparation system and markup language, and the raw data is also available for download in ZIP archive file format.
Results for different designs can be compared easily as parameters can be given as arrays. We also provide default parameter sets for some known design studies, and once processed parameters can be restored by a recovery mechanism.

If you want to work with the parameters used in this document, visit http://spacegravity.org/designer and enter code `2884-33df-8bcb'. You can also use the permalink http://spacegravity.org/designer/\#rc=2884-33df-8bcb.

The present web application was developed to quickly identify limiting noise sources common to all laser interferometric gravitational wave observatories.  Noise contributions addressed in this document are not intended to be exhaustive. Additional systems specific to the detailed observatory design might add a significant amount of excess noise. Also for most contributions white noise was assumed, however, in reality the noise shapes will be more complex.
Future updates may include additional noise contributions and individual noise shapes.

Nevertheless, the ``Gravitational Wave Observatory Designer'' is the most comprehensive simulator for a wide range of spaceborne gravitational wave detectors to our knowledge. It will educate on the subject of interferometric gravitational wave observatories, quickly show the limitations of new ideas and concepts, and help to explore the parameter space in preparation for the planned call for mission concepts for ESA's L3 mission opportunity, expected in 2016 \cite{esaCalls}.

\section*{References}
\bibliographystyle{iopart-num}
\bibliography{TheGravitationalWaveObservatoryDesigner}

\providecommand{\newblock}{}
\begin{thebibliography}{10}
\expandafter\ifx\csname url\endcsname\relax
  \def\url#1{{\tt #1}}\fi
\expandafter\ifx\csname urlprefix\endcsname\relax\def\urlprefix{URL }\fi
\providecommand{\eprint}[2][]{\url{#2}}

\bibitem{einstein1937gravitational}
Einstein A and Rosen N 1937 {\em Journal of the Franklin Institute\/} {\bf 223}
  43--54

\bibitem{taylor1989further}
Taylor J~H and Weisberg J~M 1989 {\em The Astrophysical Journal\/} {\bf 345}
  434--450

\bibitem{ade2014bicep2}
Ade P, Aikin R, Barkats D, Benton S, Bischoff C, Bock J, Brevik J, Buder I,
  Bullock E, Dowell C {\em et~al.\/} 2014 {\em arXiv preprint
  arXiv:1403.3985\/}

\bibitem{mortonson2014joint}
Mortonson M~J and Seljak U 2014 {\em arXiv preprint arXiv:1405.5857\/}

\bibitem{0264-9381-27-8-084013}
Hobbs G, Archibald A, Arzoumanian Z, Backer D, Bailes M, Bhat N~D~R, Burgay M,
  Burke-Spolaor S, Champion D, Cognard I, Coles W, Cordes J, Demorest P,
  Desvignes G, Ferdman R~D, Finn L, Freire P, Gonzalez M, Hessels J, Hotan A,
  Janssen G, Jenet F, Jessner A, Jordan C, Kaspi V, Kramer M, Kondratiev V,
  Lazio J, Lazaridis K, Lee K~J, Levin Y, Lommen A, Lorimer D, Lynch R, Lyne A,
  Manchester R, McLaughlin M, Nice D, Oslowski S, Pilia M, Possenti A, Purver
  M, Ransom S, Reynolds J, Sanidas S, Sarkissian J, Sesana A, Shannon R,
  Siemens X, Stairs I, Stappers B, Stinebring D, Theureau G, van Haasteren R,
  van Straten W, Verbiest J~P~W, Yardley D~R~B and You X~P 2010 {\em Classical
  and Quantum Gravity\/} {\bf 27} 084013
  \urlprefix\url{http://stacks.iop.org/0264-9381/27/i=8/a=084013}

\bibitem{harry2010advanced}
Harry G~M, Collaboration L~S {\em et~al.\/} 2010 {\em Classical and Quantum
  Gravity\/} {\bf 27} 084006

\bibitem{accadia2011status}
Accadia T, Acernese F, Antonucci F, Astone P, Ballardin G, Barone F, Barsuglia
  M, Basti A, Bauer T~S, Bebronne M {\em et~al.\/} 2011 {\em Classical and
  Quantum Gravity\/} {\bf 28} 114002

\bibitem{somiya2012detector}
Somiya K 2012 {\em Classical and Quantum Gravity\/} {\bf 29} 124007

\bibitem{grote2008status}
Grote H, Collaboration L~S {\em et~al.\/} 2008 {\em Classical and Quantum
  Gravity\/} {\bf 25} 114043

\bibitem{seoane2013gravitational}
Seoane P~A, Aoudia S, Audley H, Auger G, Babak S, Baker J, Barausse E, Barke S,
  Bassan M, Beckmann V {\em et~al.\/} 2013 {\em arXiv preprint
  arXiv:1305.5720\/}

\bibitem{danzmann2011lisa}
Danzmann K, Prince T {\em et~al.\/} 2011 {LISA assessment study report (Yellow
  Book)} Tech. rep.

\bibitem{jenrich2012ngo}
Jenrich O, eLISA/NGO Collaboration {\em et~al.\/} 2012 {\em SRE (2011)\/} {\bf
  19}

\bibitem{wang2013octahedron}
Wang Y, Keitel D, Babak S, Petiteau A, Otto M, Barke S, Kawazoe F, Khalaidovski
  A, M{\"u}ller V, Sch{\"u}tze D {\em et~al.\/} 2013 {\em Physical Review D\/}
  {\bf 88} 104021

\bibitem{SBarkePhD}
Barke S 2014, In preparation {\em {Inter Spacecraft Frequency Distribution for
  Future Gravitational Wave Observatories}\/} Ph.D. thesis Leibniz Universität
  Hannover Institute for Gravitational Physics

\bibitem{heinzel2011auxiliary}
Heinzel G, Esteban J~J, Barke S, Otto M, Wang Y, Garcia A~F and Danzmann K 2011
  {\em Classical and Quantum Gravity\/} {\bf 28} 094008

\bibitem{holmes1970axis}
Holmes D, Avizonis P and Wrolstad K 1970 {\em Applied optics\/} {\bf 9}
  2179--2180

\bibitem{jennrich2009lisa}
Jennrich O 2009 {\em Classical and Quantum Gravity\/} {\bf 26} 153001

\bibitem{meers1991modulation}
Meers B~J and Strain K~A 1991 {\em Physical Review A\/} {\bf 44} 4693

\bibitem{niebauer1991nonstationary}
Niebauer T, Schilling R, Danzmann K, R{\"u}diger A and Winkler W 1991 {\em
  Physical Review A\/} {\bf 43} 5022

\bibitem{esteban2011experimental}
Esteban J~J, Garc{\'\i}a A~F, Barke S, Peinado A~M, Cervantes F~G, Bykov I,
  Heinzel G and Danzmann K 2011 {\em Optics express\/} {\bf 19} 15937--15946

\bibitem{preston2010stability}
Preston A 2010 {\em {Stability of materials for use in space-based
  interferometric missions}\/}

\bibitem{schuster2014vanishing}
Schuster S, Wanner G, Tr{\"o}bs M and Heinzel G 2014 {\em arXiv preprint
  arXiv:1406.5367\/}

\bibitem{anza2005ltp}
Anza S, Armano M, Balaguer E, Benedetti M, Boatella C, Bosetti P, Bortoluzzi D,
  Brandt N, Braxmaier C, Caldwell M {\em et~al.\/} 2005 {\em Classical and
  Quantum Gravity\/} {\bf 22} S125

\bibitem{otto2012tdi}
Otto M, Heinzel G and Danzmann K 2012 {\em Classical and Quantum Gravity\/}
  {\bf 29} 205003

\bibitem{tinto2014time}
Tinto M and Dhurandhar S~V 2014 {\em Living Reviews in Relativity\/} {\bf 17}

\bibitem{esteban2009optical}
Esteban J~J, Bykov I, Mar{\'\i}n A~F~G, Heinzel G and Danzmann K 2009 {Optical
  ranging and data transfer development for LISA} {\em Journal of Physics:
  Conference Series\/} vol 154 (IOP Publishing) p 012025

\bibitem{wang2014first}
Wang Y, Heinzel G and Danzmann K 2014 {\em arXiv preprint arXiv:1402.6222\/}

\bibitem{FinalReport}
Barke S, Brause N, Bykov I, Esteban~Delgado J~J, Enggaard A, Gerberding O,
  Heinzel G, Kullmann J, Møller~Pedersen S and Rasmussen T 2014 {\em {LISA
  Metrology Systen - Final Report}\/} ({DTU Space / AEI Hannover / Axcon ApS})
  \url{http://hdl.handle.net/11858/00-001M-0000-0023-E266-6}

\bibitem{doherty1982economic}
Doherty W~J and Thadhani A~J 1982 {\em IBM Report\/}

\bibitem{tinto2002time}
Tinto M, Estabrook F~B and Armstrong J 2002 {\em Physical Review D\/} {\bf 65}
  082003

\bibitem{lrr-2014-6}
Tinto M and Dhurandhar S~V 2014 {\em Living Reviews in Relativity\/} {\bf 17}
  \urlprefix\url{http://www.livingreviews.org/lrr-2014-6}

\bibitem{cornish2001detecting}
Cornish N~J 2001 {\em Physical Review D\/} {\bf 65} 022004

\bibitem{chen2006interferometers}
Chen Y, Pai A, Somiya K, Kawamura S, Sato S, Kokeyama K, Ward R~L, Goda K and
  Mikhailov E~E 2006 {\em Physical review letters\/} {\bf 97} 151103

\bibitem{antoine}
Petiteau A 2013 private communication

\bibitem{powell2006atlas}
Powell R 2006 {An Atlas of the Universe} Website:
  \url{www.atlasoftheuniverse.com/superc.html}

\bibitem{alberto}
Sesana A 2013 private communication

\bibitem{horch2014most}
Horch E~P, Howell S~B, Everett M~E and Ciardi D~R 2014 {\em arXiv preprint
  arXiv:1409.1249\/}

\bibitem{LISAwiki}
Nelemans G {LISA wiki (verification binaries)}
  \url{http://www.astro.ru.nl/~nelemans/dokuwiki/doku.php?id=lisa_wiki}
  accessed: 2014-09-01

\bibitem{LectureA}
Miller C 2008 {Lecture 25: Grav waves from binaries}
  \url{http://www.astro.umd.edu/~miller/teaching/astr498/}

\bibitem{esaCalls}
{European Space Agency} 2014 {Timeline for Selection of L-Class Missions}
  Website: \url{http://sci.esa.int/cosmic-vision/42369-l-class-timeline/}

\end{thebibliography}

\end{document}